\documentclass[a4paper,11pt]{article}
\pdfoutput=1 % if your are submitting a pdflatex (i.e. if you have
\usepackage{jcappub} % for details on the use of the package, please
\usepackage[T1]{fontenc} % if needed

\usepackage{xcolor}
\usepackage{mathrsfs}
\usepackage{subfig}
\usepackage{romannum}

\title{\boldmath Quantized Vortices in Superfluid Dark Matter}

\author[a,1]{R. Mauland,\note{Corresponding author.}}
\author[a]{and \O. Elgar\o y}
\affiliation[a]{Institute of Theoretical Astrophysics (ITA), University of Oslo,\\
             Postboks 1029, 0315 Oslo, Norway}
\emailAdd{renate.mauland-hus@astro.uio.no}
\emailAdd{oystein.elgaroy@astro.uio.no}
\abstract{In 2015 Berezhiani \& Khoury proposed a Superfluid Dark Matter (SFDM) model where dark matter condenses and forms a superfluid on galactic scales. In the superfluid state phonons interact with baryons, resulting in a behavior similar to that of Modified Newtonian Dynamics (MOND). If one assumes that the DM condensate rotates along with the galaxy, a grid of vortices should form throughout the superfluid component if the rotation is fast enough. We aim to investigate the size and impact of the vortices on surrounding baryons, and to further investigate the parameter space of the model. We also look for a possible vortex solution of the Lagrangian presented for the SFDM theory. We first take a simple approach and investigate vortex properties in a constant density DM halo, applying knowledge from condensed matter physics. We then use the zero-temperature condensate density profile as a template to vary the DM particle mass and the energy scale, $\Lambda$, of the SFDM model. Further, we attempt to find a vortex solution of the theory by extracting the Euler-Lagrange equation with respect to the modulus of the condensate wavefunction from the full relativistic SFDM Lagrangian. For the constant density approach we find that the vortices are on millimeter scale, and separated by distances $\sim0.002\,\rm{AU}$. The parameter space of the model is found to be substantial and a reduction in the DM particle mass leads to larger vortices with a higher energy. However, none of the parameter combinations explored here give both realistic values of $\Lambda$ and vortices energetic enough to have an observational impact on the galaxy as a whole. The vortex equation extracted from the Lagrangian of the model is unstable, and no solution exhibiting the standard properties of a vortex solution is found.}

\begin{document}
\maketitle
\flushbottom

\section{Introduction}
More than eighty years after its existence was first suggested \cite{zwicky1933,zwicky1937,zwicky1933r,bertone}, little is known about the dark matter except that it probably exists. Its nature is made difficult to understand by the fact that it both has to interact very weakly with the baryonic matter and at the same time explain correlations observed on galactic scales like the Baryonic Tully-Fisher (BTF) relation. The phenomenological theory of Modified Newtonian Dynamics can explain these correlations \cite{milgrom1983}, but it has its own, quite severe problems with observations on larger scales which seem to require some form of particle dark matter \cite{famaey,Famaey2012}. 

Superfluid Dark Matter (SFDM) was proposed by Berezhiani and Khoury in \cite{original} as a way of unifying the best aspects of particle dark matter and MOND. They took the dark matter to be some axion-like light scalar which can undergo Bose-Einstein condensation and form a superfluid state at galactic scales, but remain in an ordinary gas phase at large scales. In the superfluid state the low-energy sound wave-like excitations (phonons) are assumed to interact with the baryons, and this force gives rise to a MOND-like behavior. This model can account for galactic rotation curves and correlations like the BTF relation. This idea, which can be realized in more than one way, has been explored in several papers \cite{original,Berezhiani2017}, and potential problems with the idea have also been pointed out. For example, the model proposed in the original paper \cite{original} seems to predict too large vertical accelerations for the baryonic matter in galaxies \cite{lisanti}, the double role of the phonon field introduces some tension in the model \cite{mistele2021} and some problematic acausal behaviors have been identified \cite{hertzberg2021}.

Spiral galaxies rotate, and it seems natural to assume that if the SFDM idea is true, the superfluid in galaxies will rotate, too. On very general grounds a superfluid will respond to rotation by forming a lattice of vortices with quantized circulations when spun above a critical angular speed, and the vortex lattice makes the superfluid mimic solid-body rotation (\cite{pethick_smith_2008}, Ch.9). If the same thing happens in SFDM, the assumed interaction between the baryons and the excitations of the superfluid could have an impact on the galactic rotation curve, possibly leading to tension with observations. This is the motivation for the following investigation in which we consider the properties of vortices in SFDM. While our main conclusion is that they almost certainly do not play any important dynamical role, we have in the process of reaching this conclusion found other interesting results: The available parameter space in SFDM seems to be much larger than \cite{original} and \cite{Berezhiani2017} gave the impression of. Furthermore, the Lagrangian for the model proposed in \cite{original} has one potentially problematic aspect.

The structure of this paper is as follows: In Section \ref{sec:two} we introduce a simple, constant density DM halo as a first approach to studying vortices in the SFDM model. In Section \ref{sec:parvar} we explore the parameter space of the model introduced in \cite{original} and \cite{Berezhiani2017}, paying special attention to various constraints introduced in the aforementioned papers. We then move on to searching for a vortex solution of the SFDM Lagrangian in Section \ref{sec:SFDML}, before concluding in Section \ref{sec:conc}.

%---------------------------- A simple model ----------------------------------------
\section{A Simple Model}
\label{sec:two}
 In this section we study vortices within a simple galactic model, consisting of a spherical dark matter halo with constant density. As a result, the gravitational potential will be proportional to the radius squared, enabling the use of the superfluid formalism derived for a gas trapped in a harmonic oscillator (HO) potential in \cite{pethick_smith_2008}.  The gravitational potential has the form 
 \begin{equation}
 \Phi = \frac{2}{3}\pi\rm{G}\rho r^2,
 \end{equation}
 where G is the Newtonian gravitational constant, $\rho$ is the density and $r$ is the radius from the center of the galaxy. The gravitational potential energy is then given as $V_{\Phi} = m\Phi$, which may be compared to the HO potential energy,
 \begin{equation}
 V_{\rm{HO}} = \frac{1}{2}m\omega^2r^2.
 \end{equation}
 Here $\omega$ is the angular frequency of the HO trap, $m$ is the mass of the trapped particles and $r$ is the radius from the center. This leads to the relation
 \begin{equation}
 \omega = \sqrt{\frac{4\pi\rm{G}\rho}{3}}\label{eq:omega}
 \end{equation}
 for the density and angular frequency. Using this relation, we may apply the equations derived in Ch. 9 of \cite{pethick_smith_2008} for a rotating superfluid in a HO trap to describe our galaxy. 
 
 %----------------------------- Vortex Properties ---------------------------------------
 \subsection{Vortex Properties}
 \label{sec:properties}
\begin{figure}
   \centering
   %\vfill
   \includegraphics[scale=1.0]{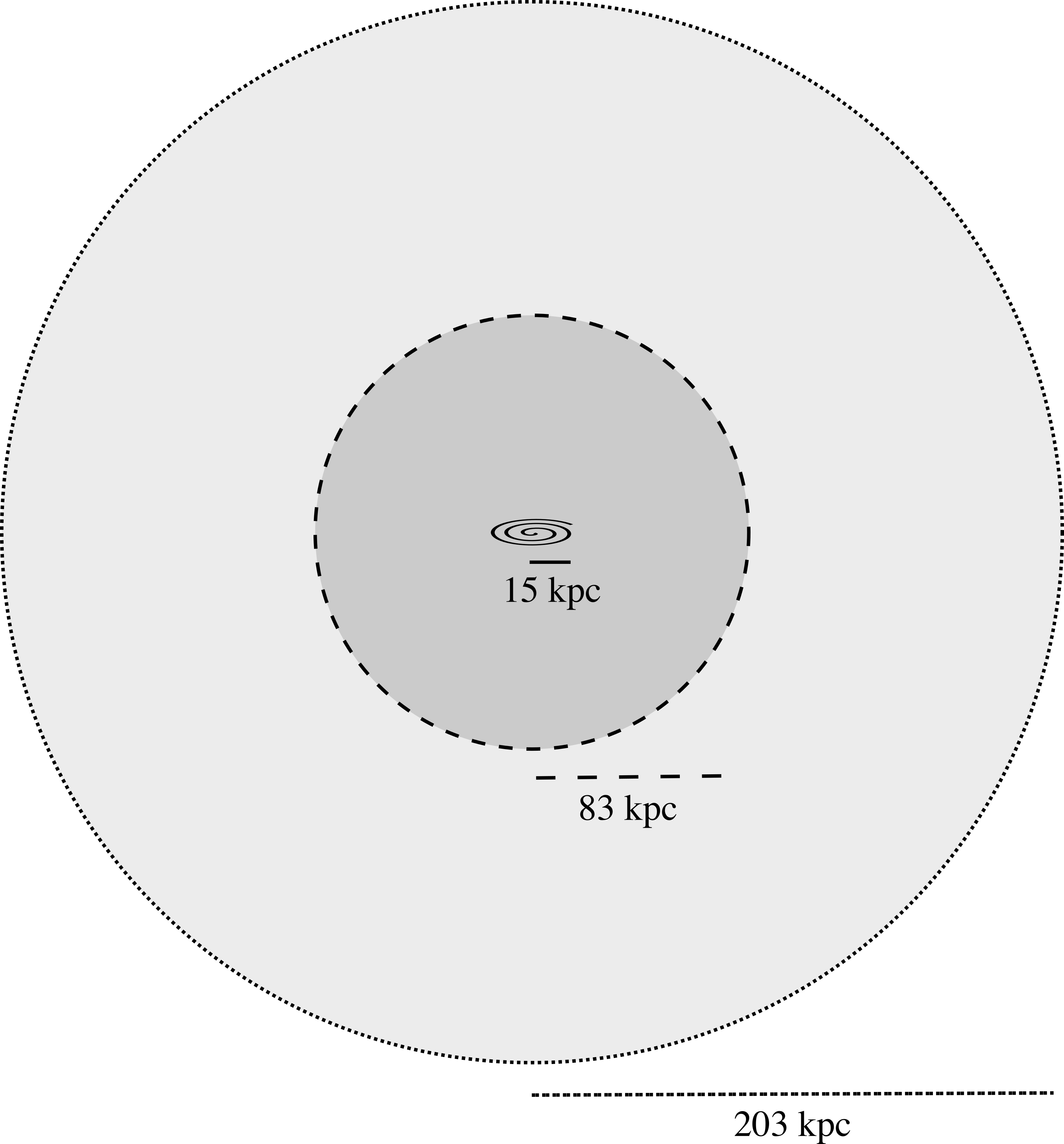}
      \caption{Extent of the superfluid component compared to the size of the dark matter halo. The \textit{full} line represents the radius of the galactic disk, the \textit{dashed} line the radius of the superfluid dark matter component and the \textit{dotted} line the full extent of the dark matter halo. Beyond the superfluid radius the dark matter behaves as regular CDM particles, as in the $\Lambda$CDM model.
              }
         \label{fig:halocartoon}
   \end{figure}

 A well known property of a rotating superfluid is the formation of vortices. As galaxies rotate, vortices should form in the superfluid dark matter component, assuming that the dark matter halo rotates as well. This could have an impact on the surrounding baryonic matter and might even lead to an observational constraint for the SFDM model. We will here calculate the following vortex properties:
 \begin{enumerate}
 \item The energy of a vortex.
 \item The coherence length of the superfluid.
 \item The critical angular velocity.
 \item The superfluid/condensate radius.
 \item The intervortex spacing.
 \item The ``mass'' of a single vortex.
 \end{enumerate}
From \cite{pethick_smith_2008}, Ch. 9, we have that the energy per unit length of a vortex (in 2D) is given as
\begin{align}
\varepsilon_{\rm{v}} \simeq \pi n(0)\frac{\hbar^2}{m}\ln\Big(0.888\frac{R}{\xi_0}\Big)\label{eq:epsilon},
\end{align}
for a Bose-Einstein condensate in a HO potential. The potential is assumed to be rotationally invariant about the $z$-axis and the vortex is centered at the rotational axis. This equation is valid as long as there are enough particles that the Thomas-Fermi approximation holds, which we assume to be the case\footnote{The Thomas-Fermi approximation states that a high enough number of particles will allow us to ignore the kinetic energy term in the Gross-Pitaevskii equation, from which this equation is derived (see Ch.6, \cite{pethick_smith_2008})}. Here $n(0)$ is the number density in the center of the condensate, in the absence of a vortex, $m$ is the DM particle mass and $R$ is the radius of the condensate. Finally, $\xi_0$ is the coherence length of the superfluid, which also describes the size of the vortex cores. From \cite{pethick_smith_2008} it is given as
\begin{align}
\xi_0 = \frac{\hbar}{m\omega R}, \label{eq:coherence}
\end{align}
for a condensate trapped in a HO potential. Here $\omega$ is defined by equation (\ref{eq:omega}). %The coherence length is important as it determines the size of the vortex core.
The critical angular velocity for vortex formation in a HO trap is, from \cite{pethick_smith_2008},
\begin{align}
\Omega_c = \frac{5}{2}\frac{\hbar}{mR^2}\ln\Big(0.671\frac{R}{\xi_0}\Big)\label{eq:omega_c},
\end{align}
with the variables as previously described. Equations (\ref{eq:epsilon}), (\ref{eq:coherence}) and (\ref{eq:omega_c}) all depend on the radius of the superfluid component, $R$, which from \cite{Berezhiani2017} is given as the thermalization radius
\begin{align}
R_T \lesssim 310\,\rm{kpc}\,\Big(\frac{m}{\rm{eV}}\Big)^{-8/7}\Big(\frac{M}{10^{12}M_{\odot}}\Big)^{1/7}\Big(\frac{\sigma/m}{\rm{cm}^2/\rm{g}}\Big)^{2/7}\label{eq:R_T}.
\end{align}
Here $m$ is the DM particle mass, $M$ is the mass of the galaxy and $\sigma/m$ is the self-interaction cross section per unit mass of the DM particles. Based on \cite{Berezhiani2017} we will use 
\begin{align}
m = 1\,\rm{eV}\qquad\rm{and}\qquad\sigma/m=0.01\,\rm{cm}^2/\rm{g}\label{eq:variables}.
\end{align}
For the constant density of our DM sphere, we will use a  density of 200 times the present day critical density, 
\begin{align}
\rho_{200} = 200\rho_{\rm{crit}}=200\frac{3H_0^2}{8\pi G}\simeq 1.84\times10^{-27}\,\rm{g}/\rm{cm}^3.\label{eq:crit_density}
\end{align}
Here $H_0 = 70\,\rm{km}\rm{s}^{-1}\rm{Mpc}^{-1}$ and $G$ is the Newtonian gravitational constant. Using the mass $M = 10^{12}M_{\odot}$ for a Milky Way-sized galaxy, we obtain the following values for points 1-4 in our list
\begin{align}
%\rho &= 1.84\times10^{-27}\,\rm{g}/\rm{cm}^3\label{eq:20}\\
1.\;\;\varepsilon_{\rm{v}} &\simeq 1.135\times10^{-13}\, \rm{g}\,\rm{cm}\;\rm{s}^{-2}\label{eq:23},\\
2. \;\;\xi_0 &\simeq 0.1016\;\rm{cm}\label{eq:xi_val},\\
3. \;\Omega_c &\simeq 1.253\times10^{-39}\;\rm{s}^{-1}\label{eq:crit},\\
4. \;R_T &\lesssim 83.16\;\rm{kpc}\label{eq:RT_val}.
\end{align}
From \cite{Berezhiani2017}, the virial radius is given as 
\begin{align}
R_{200} \simeq 203\Big(\frac{M}{10^{12}M_{\odot}}\Big)^{1/3}\;\rm{kpc} = 203 \;\rm{kpc}\label{eq:virial}
\end{align}
for a Milky Way-sized galaxy. The relative proportions between the galactic disk, the superfluid component and the full DM halo is roughly illustrated in Figure \ref{fig:halocartoon}.

%-------------------------------------- Two column figure (place early!)
   %\begin{figure*}
   %\centering
   %\includegraphics{halo_composition3.png}
   %%%\includegraphics{empty.eps}
   %%%\includegraphics{empty.eps}
   %\caption{Extent of the superfluid component compared to the size of the dark matter halo. The \textit{full} line represents the radius of the galactic disk, the \textit{dashed} line the radius of the superfluid dark matter component and the \textit{dotted} line the full extent of the dark matter halo. Beyond the superfluid radius the dark matter behaves as regular CDM particles, as in the $\Lambda$CDM model.}
              %\label{FigGam}%
   % \end{figure*}
%

The typical angular velocity of a Milky Way-sized galaxy is
\begin{align}
\Omega_{\rm{MW}} = \frac{v}{r}=\frac{2\pi r/T}{r}=\frac{2\pi}{T}=\frac{2\pi}{250\,\rm{Myr}}\approx 8\times10^{-16}\,\rm{s}^{-1}.\label{eq:ang_vel}
\end{align}
The value found for the critical velocity in equation (\ref{eq:crit}) is less than the value given in equation (\ref{eq:ang_vel}), showing that vortices should form in the rotating superfluid component of the DM halo. As the angular velocity in a condensate is increased beyond the critical value, several vortices will form in a grid, each with a single quantum of circulation. Assuming uniform rotation, we may use the results from \cite{pethick_smith_2008} relating the number of vortices per unit area to the angular velocity,
\begin{align}
    n_{v} = \frac{2m\Omega}{h}=\frac{1}{\pi a^2_{\Omega}}. \label{eq:intervortex}
\end{align}
Here $a_{\Omega}=\sqrt{\hbar/m\Omega}$ is the intervortex spacing. For a Milky Way-sized galaxy this results in
\begin{align}
    a_{\Omega}\sim 2.7\times10^{10}\,\rm{cm}\approx 0.002\,\rm{AU}.
\end{align}

The last remaining vortex quantity is the ``mass'' of a single vortex. The energy calculated previously was only two dimensional, but \cite{pethick_smith_2008} also provides an expression for the three dimensional energy of a single vortex sitting at the rotational axis of the system, 
\begin{align}
    E = \frac{4\pi n(0,0)}{3}\frac{\hbar^2}{m}Z\ln\Big(0.671\frac{R}{\xi_0}\Big).\label{eq:vortex_energy}
\end{align}
Here $n(0,0)$ is the number density in the center in absence of a vortex, $Z$ is the height of the condensate (meaning that the shape is simplified to be cylindrical) and $R$ is the radius of the condensate. Inserting the values from equations (\ref{eq:crit_density}), (\ref{eq:xi_val}) and (\ref{eq:RT_val}) gives
\begin{align}
    E \approx 3863\,\rm{J},
\end{align}
corresponding to a mass of
\begin{align}
    m_{\rm{vortex}}\sim 4\times 10^{-14}\,\rm{kg}.
\end{align}
Based on a condensate radius of $83\,\rm{kpc}$ and a vortex separation of $0.002\,\rm{AU}$, there is a total of $\sim10^{26}$ vortices within the condensate, amounting to a mass of $\sim10^{12}\,\rm{kg}$.

%----------------------------- Critical Temperature ---------------------------------------
\subsection{Critical Temperature}
\label{sec:temp}
One of the main criteria presented in \cite{original} is that the dark matter condenses and forms a superfluid. In this section we will calculate the critical temperature of the dark matter and the temperature of the dark matter as a result of its self-interaction. 

From \cite{pethick_smith_2008} the critical temperature in a condensate trapped in a HO potential is given as
\begin{align}
    T_c \approx 4.5\Big(\frac{\bar{f}}{100\,\rm{Hz}}N^{1/3}\Big),\label{eq:Tc}
\end{align}
where $N$ is the number of (dark matter) particles and
\begin{align}
    \bar{f} = \bar{\omega}/2\pi,\quad \bar{\omega} = (\omega_x\omega_y\omega_z)^{1/3}.
\end{align}
From the density and volume of the full DM halo, the number of dark matter particles is
\begin{align}
    N\approx 1.1\times10^{78}.\label{eq:N}
\end{align}
Using the angular velocity of equation (\ref{eq:omega}) together with equations (\ref{eq:Tc}) and (\ref{eq:N}) gives the critical temperature
\begin{align}
    T_c \approx 17\,\rm{mK}
\end{align}
for the DM halo. 

In the SFDM model presented in \cite{original}, the dark matter particles self-interact, and we may use the virial theorem to translate gravitational energy into thermal energy,
\begin{align}
    \frac{3}{5}\frac{GM}{R} = \frac{3}{2}\frac{k_BT}{m}.
\end{align}
Here $M$ is the mass of the halo, $R$ is the radius and $m$ is the mass of a single DM particle. Inserting
\begin{align}
    M=10^{12}M_{\odot},\quad R=203\,\rm{kpc}\quad \rm{and}\quad m=1\,\rm{eV}
\end{align}
we get a temperature of
\begin{align}
    T\approx 1.1\,\rm{mK}.
\end{align}
This shows that the DM has a temperature below the critical temperature inside the halo, meaning that it would condense and form a superfluid\footnote{In \cite{original} they instead use the overlapping of the de Broglie wavelength, $\lambda_{dB}\sim1/mv$, of the particles to decide if the DM will condense into a superfluid.}.

%-------------------------- Parameter Variation ------------------------------------------
\section{Parameter Variation}
\label{sec:parvar}
In \cite{original} and \cite{Berezhiani2017}, the SFDM model is presented as
\begin{align}
\mathscr{L} = \frac{2\Lambda(2m^{3/2})}{3}X\sqrt{|X - \beta Y|}-\alpha\frac{\Lambda}{M_{\rm{Pl}}}\theta\rho_b.\label{eq:modelll}
\end{align}
Here $X = \mu -m\Phi+\dot{\theta}-(\vec{\nabla}\theta)^2/2m$, where $\theta$ is the phonon scalar field, $\Phi$ is the gravitational potential and $\mu$ is the chemical potential. $Y$ is the scalar product of the normal and superfluid velocity fields, and $\beta$ parametrises finite-temperature effects. $M_{\rm{Pl}}$ is the reduced Planck mass, $\alpha$ is a coupling constant and $\rho_b$ is the density profile of the baryons. The first term has the form of the MOND scalar action ($\mathscr{L}_{\rm{MOND}}\sim\Lambda X\sqrt{|X|}$), while the second term describes the interaction between phonons and baryons. In this section we are interested in $m$, the dark matter particle mass, and $\Lambda$, the energy scale that defines the validity range of the model. In the previous section we considered $m=1\,\rm{eV}$ and $\sigma/m = 0.01\,\rm{cm}^2/\rm{g}$ as fiducial values. This is based on \cite{Berezhiani2017}, where the value $\Lambda=0.05\,\rm{meV}$ was also used. In \cite{original}, the values $m=0.6\,\rm{eV}$ and $\Lambda=0.2\,\rm{meV}$ were chosen. In our calculations we will use the values $m=1\,\rm{eV}$, $\sigma/m = 0.01\,\rm{cm}^2/\rm{g}$ and $\Lambda=0.2\,\rm{meV}$ as fiducial values. This is mostly based on \cite{Berezhiani2017}, as this is the most recent assessment, but also chosen to fulfill all constraints set up in the aforementioned paper, along with ensuring that the condensate radius is approximately the same in the two different approaches of \cite{original} and \cite{Berezhiani2017}. In this section we will check if it is possible to vary the values of $m$ and $\Lambda$ and still conform to all constraints on the SFDM model as presented in \cite{Berezhiani2017}.

%-------------------------------- Condensate Halo Density profile ------------------------------------
\subsection{Condensate Halo Density Profile}
\label{sec:profile}
\begin{figure}
   \centering
   \includegraphics[trim={0 0 0 1cm},clip,scale=0.5]{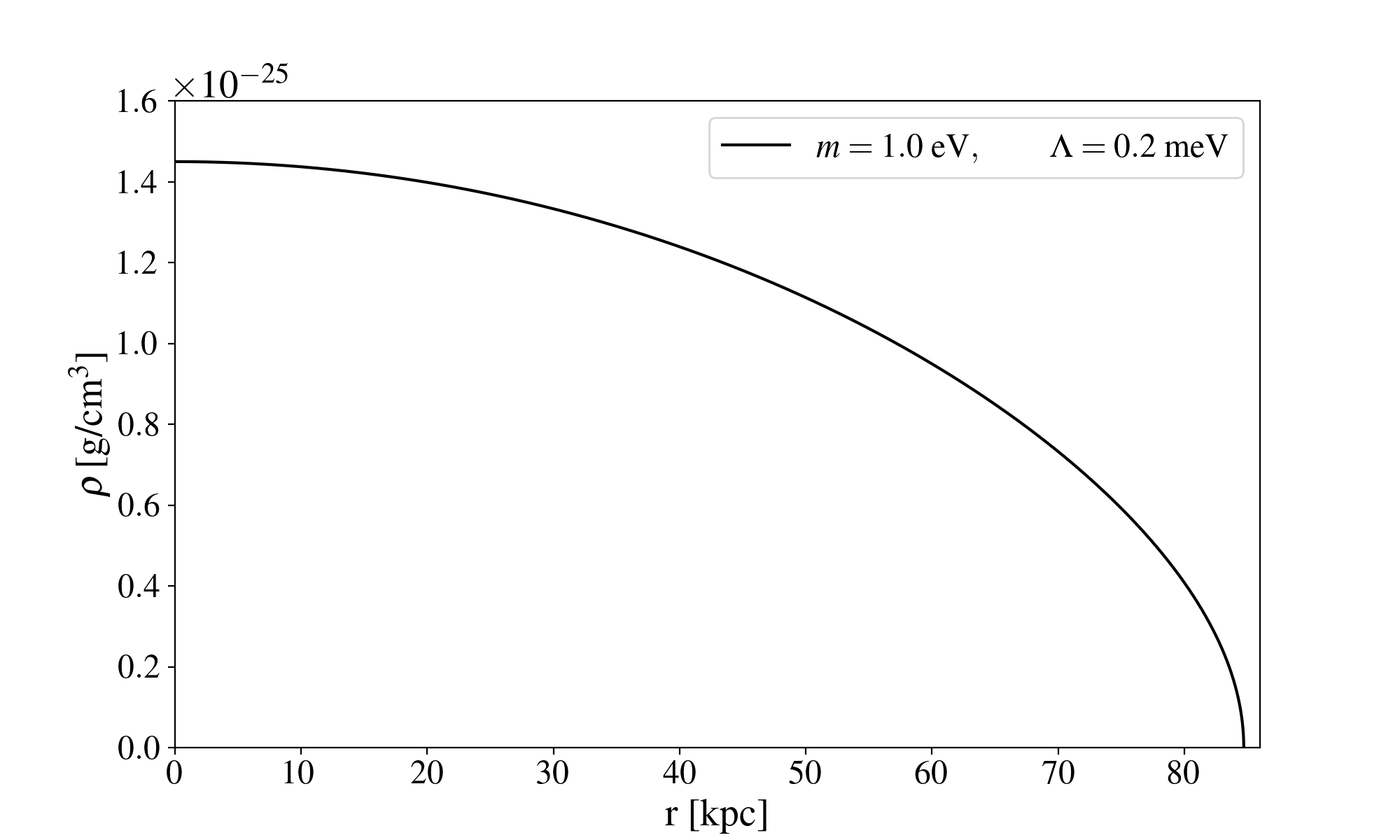}
      \caption{Zero-temperature condensate dark matter density profile, ignoring baryons. Defining the condensate radius as the point where the density profile reaches zero, we obtain a condensate radius of $\sim 84.7\;\rm{kpc}$.
              }
         \label{fig:dens_profile}
   \end{figure}

We will start by reproducing the zero temperature condensate halo density profile presented in \cite{original}\footnote{A more advanced calculation can be found in section V of \cite{Berezhiani2017}.}. Assuming a static, spherically symmetric halo we have the relation
\begin{align}
\frac{1}{\rho(r)}\frac{dP}{dr} = - \frac{d\Phi}{dr} = -\frac{Gm(r)}{r^2}\label{eq:4.1.1}
\end{align}
for the pressure and acceleration. Here $\Phi = -Gm/r$ is the regular gravitational potential field. Taking the derivative of equation (\ref{eq:4.1.1}) with respect to $r$ and replacing the mass by the expression 
\begin{align}
m(r) = \int_0^r\int_0^{4\pi}r^2\rho(r)\,d\Omega\,dr\label{eq:4.1.2}
\end{align}
results in the equation
\begin{align}
    \frac{d}{dr}\Big(\frac{1}{\rho(r)}\frac{dP}{dr} \Big) = -\frac{2}{r}\frac{1}{\rho(r)}\frac{dP}{dr} - 4\pi G\rho(r)\label{eq:4.1.8}.
\end{align}
By rearranging the terms and multiplying by $r^2$ we obtain
\begin{align}
\frac{1}{r^2}\frac{d}{dr}\Big(r^2\frac{1}{\rho(r)}\frac{dP}{dr} \Big)= -4\pi G\rho(r)\label{eq:4.1.11},
\end{align}
which is the Poisson equation in spherical coordinates (with rotational symmetry). From \cite{original} we have the equation of state
\begin{align}
P = \frac{\rho^3}{12\Lambda^2m^6}=K\rho^3\label{eq:4.1.12}
\end{align}
for the SFDM, and we may insert it into equation (\ref{eq:4.1.11}). Introducing the variables $y = (\rho/\rho_{\rm{center}})^2$ and $x = r/r_0$ where $r_0^2 = 3K\rho_{\rm{center}}/8\pi G$ and $\rho_{\rm{center}}$ is the central density of the halo, we get
\begin{equation}
\frac{1}{x^2}\frac{d}{dx}\Big(x^2\frac{dy}{dx}\Big) = -y^{1/2}.\label{eq:4.1.16}
\end{equation}
This is the Lane-Emden equation for $n=1/2$ and may be solved numerically using the initial conditions $y(0)=1$ and $y'(0)=0$. In \cite{original} the central density of the condensate halo is given as
\begin{align}
\rho_{\rm{center}} \simeq \Big(\frac{M}{10^{12}M_{\odot}}\Big)^{2/5}\Big(\frac{m}{\rm{eV}}\Big)^{18/5}\Big(\frac{\Lambda}{\rm{meV}}\Big)^{6/5}10^{-24}\,\rm{g}/\rm{cm}^3 \label{eq:central_density}.
\end{align}
Using $M=10^{12}M_{\odot}$, $m=1\,\rm{eV}$ and $\Lambda=0.2\,\rm{meV}$ we get the halo density profile shown in Figure \ref{fig:dens_profile}, resulting in a condensate radius of $\sim 84.7\,\rm{kpc}$. This is close to the constraint $R_T \lesssim 83\;\rm{kpc}$ found in section \ref{sec:two}, however, it does not have to be. Defining the condensate radius by where the condensate density hits zero was done in \cite{original}, but in \cite{Berezhiani2017} a new assessment of the dark matter density profile was performed, which led to a new definition of the condensate, namely the thermalization radius, $R_T$, used in the previous section. This new thermalization radius is in general lower than the radius obtained from the zero temperature profile, depending on the choice of $\Lambda$. This is why we have chosen $\Lambda=0.2\,\rm{meV}$. The given value ensures that the radius obtained by the methods in this section is close to the upper bound on the radius obtained from the $R_T$ expression. In the following, we will therefore assume that the two radii are the same.

\begin{figure}
   \centering
   \includegraphics[trim={0 0 0 1.5cm},clip,scale=0.5]{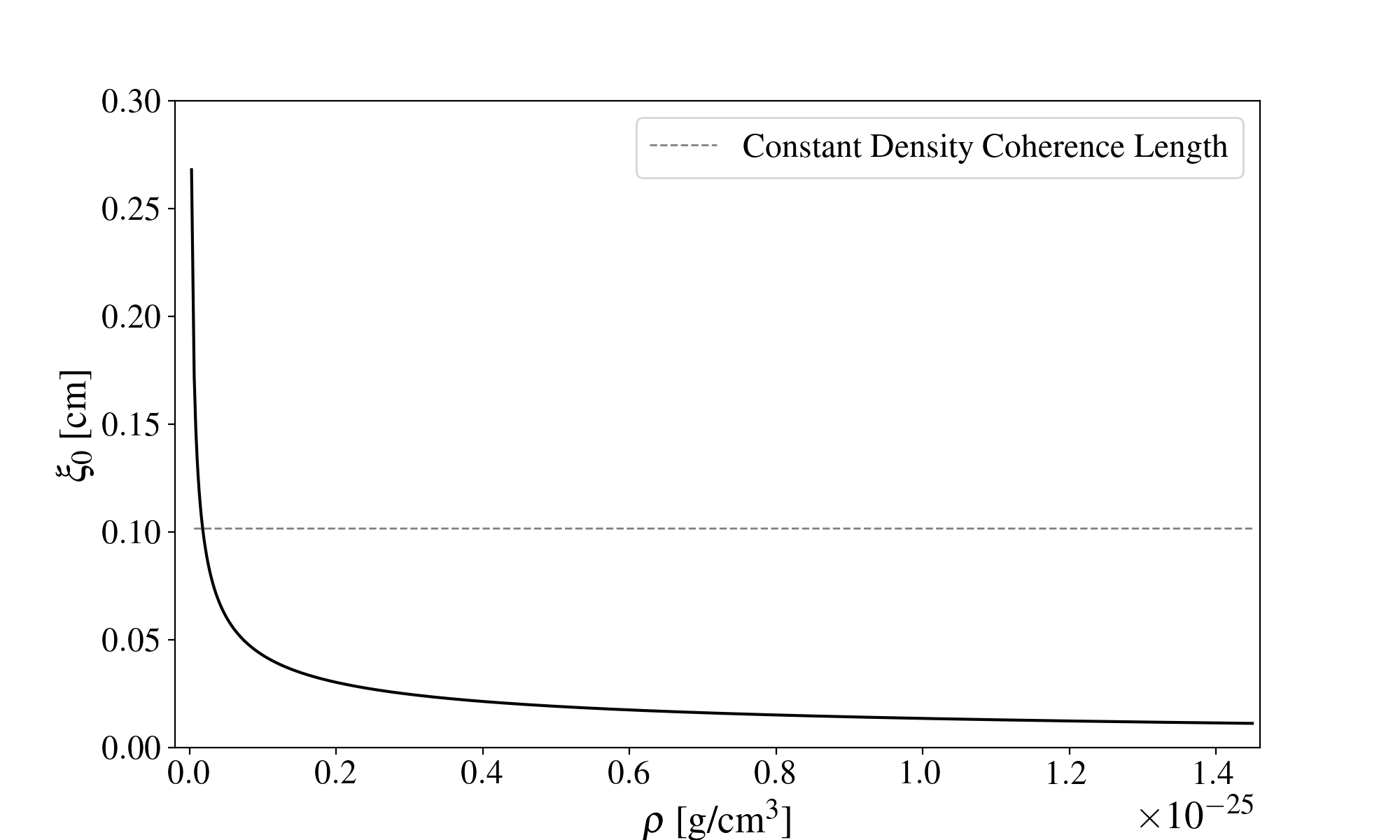}
      \caption{Coherence length behavior throughout the superfluid DM component, showing how the size of the vortex core changes as a function of density. The dashed \textit{gray} line marks the coherence length resulting from the constant density case, presented in section \ref{sec:two}, using $R_T \approx 83\,\rm{kpc}$. The edge of the condensate component is at density zero, moving towards the center as the density increases.
              }
         \label{fig:coherence_dens}
   \end{figure}

%\subsubsection{Vortex core size} 
% Dont need to put this in its own section?
An immediate consequence of the SFDM density no longer being constant, as opposed to the simple model in Section \ref{sec:two}, is that the coherence length will vary throughout the condensate. As the coherence length describes the vortex core size, this is interesting to study. A plot illustrating the behavior as a function of density is shown in Figure \ref{fig:coherence_dens}.
Here it can be seen that the coherence length (vortex core) is at its smallest near the center of the condensate, and increases towards the edge:
\begin{align}
    \xi_{\rm{center}} \approx 1.1\times10^{-4}\,\rm{m},\quad \xi_{\rm{edge}}\approx 2.7\times10^{-3}\,\rm{m}.
\end{align}
This calculation is based on expressions from Section \ref{sec:two}, which again is based on a constant condensate density. Now we have a density that varies with radius, however, on vortex scales, the density may still be assumed to stay constant.

%-------------------------------- One Meter vortex Core ------------------------------------
\subsection{One Meter Vortex Core}
So far, we have found the vortices to be small and seemingly unimportant for the dynamics of the surrounding baryons. In this section we want to address the possibility of changing the parameters of the model to obtain larger and more ``massive'' vortices. 

The coherence length equation, $\xi_0=\frac{\hbar}{m\omega R}$, shows that a reduction in the DM particle mass will increase the vortex core size, as long as the density profile, and thereby the condensate radius, remains constant. From equation (\ref{eq:vortex_energy}) a smaller DM particle mass, even though it results in a larger vortex core, will give a higher energy per vortex. The condensate density profile depends on $\rho_{\rm{center}}$, which again depends upon $m$, $\Lambda$ and $M$. $M$ is the total mass of the galaxy, and we will take this as constant. If we want to decrease $m$, and still keep the density profile constant, we must check if there are any values of $\Lambda$ which will allow this. To do so, we will go through a number of checkpoints to make sure that none of the constrains set up for the SFDM model in \cite{Berezhiani2017} are violated:
\begin{enumerate}
\item \textit{The Particle Mass:} Does the model allow a decrease of the DM particle mass, $m$?
\item \textit{The $\Lambda$-value:} Can we find a $\Lambda$-value that reproduces the original density profile? Does it still fulfill the constraints presented in \cite{Berezhiani2017}?
\item \textit{The Self-Interaction:} How does changing $m$ and $\Lambda$ affect the self-interaction cross section? Does it still fulfill the constraints presented in \cite{Berezhiani2017}?
\item \textit{The Temperature:} Is the halo temperature still below the critical temperature?
\item \textit{The Critical Angular Velocity:} Is the angular velocity of the galaxy still higher than the critical angular velocity?
\item \textit{The Energy of a Single Vortex:} Are we actually achieving a larger energy/mass per vortex, as expected?
\item \textit{The $\alpha$ Parameter:} Does the coupling constant, $\alpha$, from equation (\ref{eq:modelll}) affect our parameter changes?
\item \textit{Rotation Curves:} Do the parameter changes affect the ability of the model to reproduce galactic rotation curves?
\end{enumerate}
In order to neatly go through the first six points on the list, we will choose a specific vortex core size of one meter. The equation describing the vortex core size depends upon the density profile, which makes it clear that a mass that results in a one meter vortex core at the center does not result in a one meter vortex core at the edge, and vice versa. We will consider the case of a one meter vortex core at the center of the SFDM halo.

 %-------------------------------- The Particle Mass ------------------------------------
\subsubsection{The Particle Mass}
\label{sec:mass_test}
As the vortex core size is $1\,\rm{m}$, we may assume the dark matter density to be constant at the given length scales. We can then use equations (\ref{eq:omega}) and (\ref{eq:coherence}) from section \ref{sec:two}, which combined gives the expression
\begin{align}
    m = \frac{\sqrt{3}\hbar}{\xi_0\sqrt{4\pi G\rho}R}
\end{align}
for the DM particle mass. Inserting $R = 84.7\,\rm{kpc}$, the density at the center of the condensate and $\xi_0=1\,\rm{m}$ we arrive at the mass required to produce a one meter vortex core at the center of the condensate:
\begin{align}
m \approx 1.1\times10^{-4}\;\rm{eV}.\label{eq:4.3.1}
\end{align}
In other words, obtaining a one meter vortex core in the center requires a mass of $\sim10^{-4}\,\rm{eV}$, which again would result in larger vortices at the edge ($\sim24\,\rm{m}$). The general trend is therefore, the lower the DM particle mass, the larger the vortices.

We must now check if the model allows us to change the mass of the DM particle in this way. The radius of the condensate component should extend beyond $60\,\rm{kpc}$ for a Milky Way-like galaxy. This is to ensure that the superfluid model can explain the observed flat rotation curves of galaxies. Demanding this results in an upper bound on the DM particle mass, presented in equation (24) of \cite{Berezhiani2017}:
\begin{align}
m \lesssim 4.2\Big(\frac{\sigma/m}{\rm{cm}^2/\rm{g}}\Big)^{1/4}\,\rm{eV}.\label{eq:m_constraint}
\end{align}
For $\sigma/m=0.01\,\rm{cm}^2/\rm{g}$, this gives $m \lesssim 1.33\;\rm{eV}$, which fits the original value of $1\,\rm{eV}$. As this is an upper bound it does not rule out the one meter vortex core masses in equation (\ref{eq:4.3.1}). However, changing both $m$ and $\Lambda$ might result in changes of $\sigma/m$, and this point will therefore be revisited.

%-------------------------------- The Lambda-Value ------------------------------------
\subsubsection{The $\Lambda$-value}
\label{sec:lambda_value}
\begin{figure}
\centering
\includegraphics[trim={0 0 0 1cm},clip,scale=0.5]{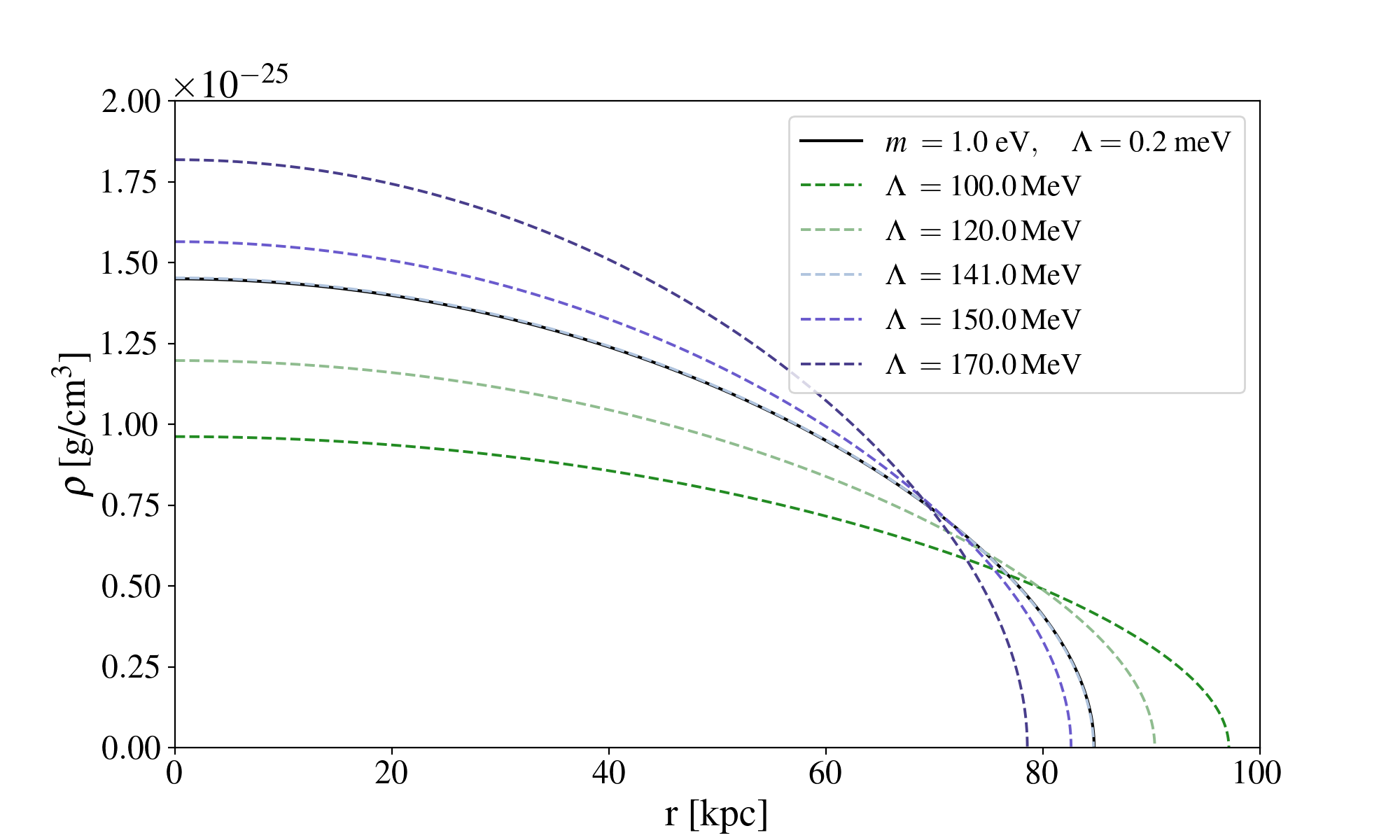}
\caption{Dark matter density profile fitting for the value $m \approx 1.1\times10^{-4}\;\rm{eV}$. A larger value of $\Lambda$ is required for the halo profile to replicate that found in Section \ref{sec:profile}.}
\label{fig:lambda_test_centeredge}
\end{figure}

Now we want to find a value for $\Lambda$ which will allow us to reproduce the halo profile in Figure \ref{fig:dens_profile}. We can again calculate the density profile, but with a new value of $m$ and $\Lambda$, and check how closely the new profile matches the old. A plot showing a few different values of $\Lambda$, together with the original density profile, is presented in Figure \ref{fig:lambda_test_centeredge} for the mass in equation (\ref{eq:4.3.1}).
From Figure \ref{fig:lambda_test_centeredge} it is evident that there does exist a $\Lambda$-value that will, in combination with the new particle mass presented in equation (\ref{eq:4.3.1}), reproduce the density profile obtained using the original values: $m=1\,\rm{eV}$, $\sigma/m=0.01\,\rm{cm}^2/\rm{g}$ and $\Lambda=0.2\,\rm{meV}$. The required $\Lambda$-value is
\begin{align}
\Lambda \approx 141\,\rm{MeV}.\label{eq:4.3.2}
\end{align}

Again we have to check that the new changes fulfill the constrains presented in \cite{Berezhiani2017}. Demanding that the SFDM model allows for formation of the lightest observed halos, such as ultra-faint dwarf satellites, \cite{Berezhiani2017} gives 
\begin{align}
    \Lambda m^3 \gtrsim 0.08 \,\rm{meV}\times\rm{eV}^3\label{eq:lambda_constraint}
\end{align}
as shown in equation (28) of the aforementioned paper. Inserting the value of $m$ given in equation (\ref{eq:4.3.1}) results in 
\begin{align}
\Lambda \gtrsim 56\;\rm{MeV},
\end{align}
which equation (\ref{eq:4.3.2}) fulfills. 

%-------------------------------- The Self-Interaction ------------------------------------
\subsubsection{The Self-Interaction}
\begin{figure}
   \centering
   \includegraphics[trim={0 0 0 1cm},clip,scale=0.5]{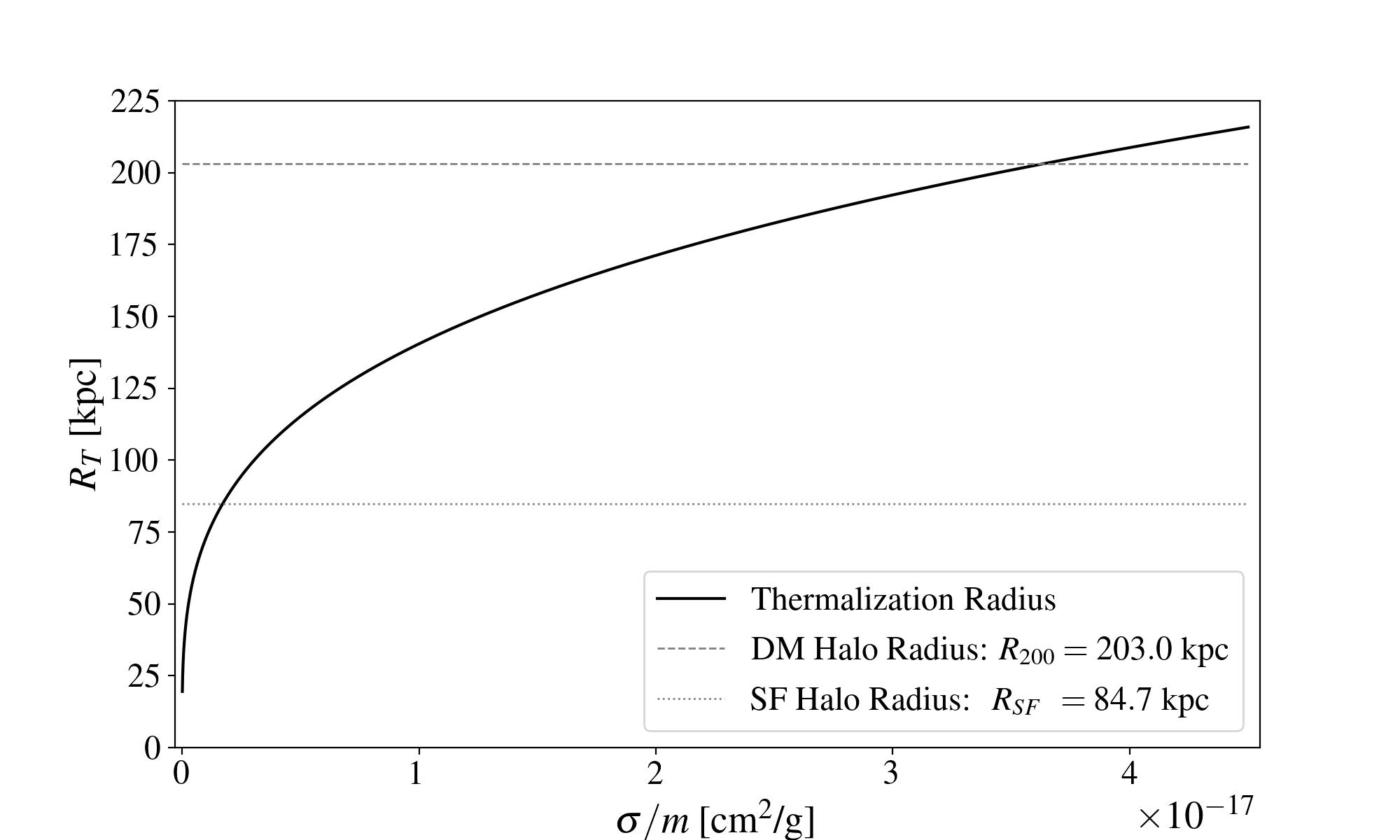}
      \caption{Cross section dependence on thermalization radius. The new $m$ and $\Lambda$ values for a one meter central vortex core require a small cross section per unit mass, $\sim10^{-18}\,\rm{cm}^2/\rm{g}$, for the thermalization radius to be $84.7\,\rm{kpc}$. As an additional comment, the superfluid halo radius displayed in this plot represents the upper limit of equation (\ref{eq:R_T}).
              }
         \label{fig:r_sigm}
   \end{figure}

Having reproduced the original density profile with new values of $m$ and $\Lambda$, we have the same condensate radius as before, $84.7\,\rm{kpc}$. In Section \ref{sec:two} we used equation (\ref{eq:R_T}) to describe the thermalization radius, within which the dark matter thermalizes and condenses into a superfluid according to \cite{Berezhiani2017}. This expression depends upon the self-interaction cross section as shown in Figure \ref{fig:r_sigm}.

Previously, we have used two different ways of defining the size of the condensate component, namely the thermalization radius (equation \ref{eq:R_T}), and the radius where the zero temperature condensate density profile hits zero. In our case, we will assume these two to be the same\footnote{This is not completely accurate as the thermalization radius is a new consideration of \cite{Berezhiani2017}, which operates with a different dark matter density profile than the zero-temperature profile considered in \cite{original}, and used by us. We have, however, chosen values of $\Lambda$ to ensure that this is fine for a simple first approach.}, which means that we now know the value of the cross section by where the dotted line intersects the black one, approximately at
\begin{align}
\sigma/m &\approx 1.7\times10^{-18}\,\rm{cm}^2/\rm{g}.\label{eq:sigm_r1}
\end{align}

\begin{figure}
   \centering
   \includegraphics[trim={0 0 0 1.5cm},clip,scale=0.5]{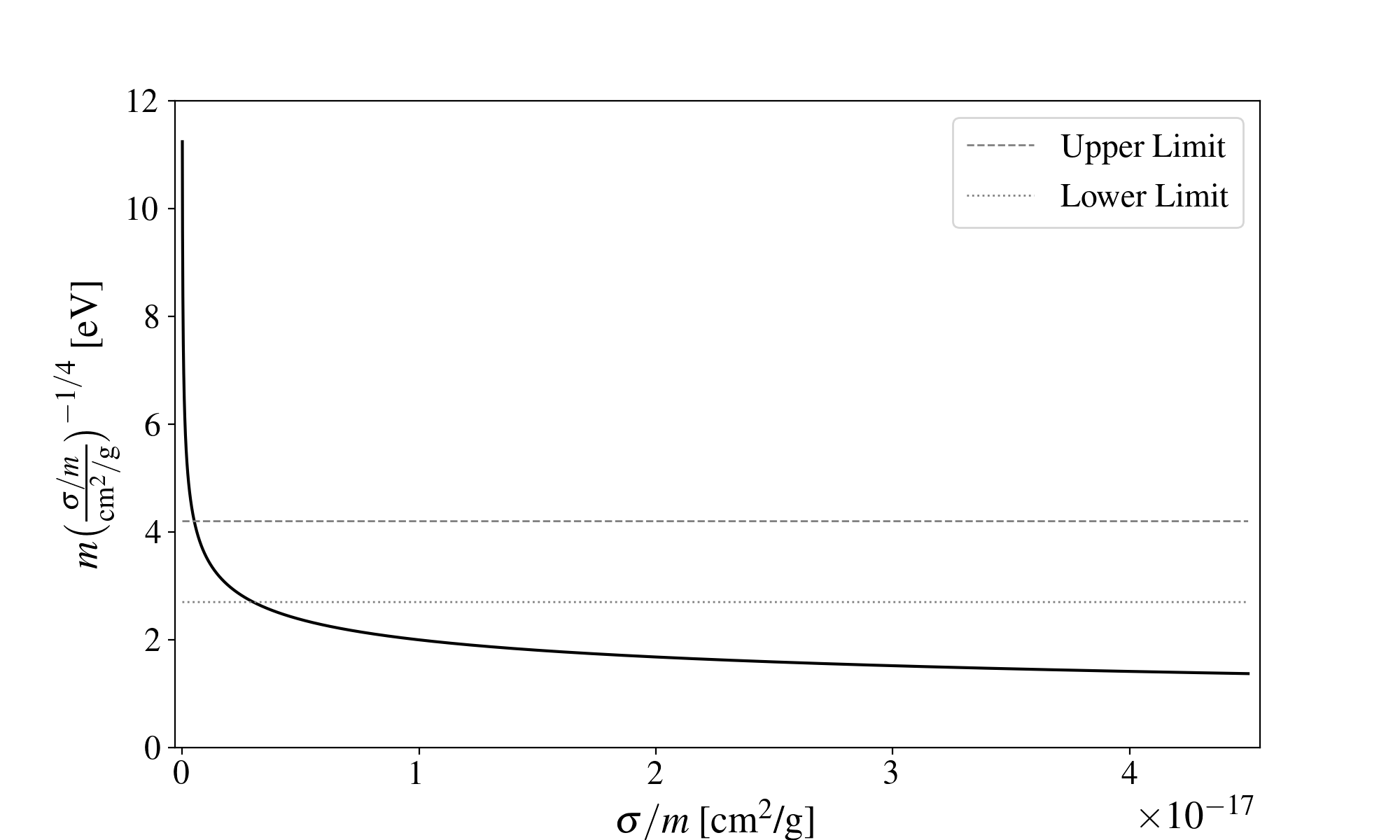}
      \caption{Visualization of the cross section constraints presented in equation (\ref{eq:m_constraint_ul}), using a dark matter particle mass of $m \simeq 1.1\times10^{-4}\,\rm{eV}$. There is only a narrow window in which $\sigma/m$ fulfills the constraints for the given DM particle mass.
              }
         \label{fig:m_sigm_ul}
   \end{figure}

The model presented in \cite{original} requires galaxies to have a significant superfluid component in order to reproduce the observations that seem to favor a MONDian behavior. Galaxy clusters, on the contrary, should not have a large superfluid component, as we want them to behave more closely to the predictions of the $\Lambda$CDM model. Altogether, this results in a lower bound on the DM particle mass. In combination with previous constraints (equation \ref{eq:m_constraint}) this amounts to 
\begin{align}
2.7\,\rm{eV}\lesssim m\Big(\frac{\sigma/m}{\rm{cm}^2/\rm{g}}\Big)^{-1/4}\lesssim 4.2\,\rm{eV}, \label{eq:m_constraint_ul}
\end{align}
which may also be seen as a condition for the cross section, if the mass is already known. For our one meter vortex core center mass, this relation is plotted in Figure \ref{fig:m_sigm_ul}.
As illustrated in Figure \ref{fig:m_sigm_ul}, there is a small window for the value of the cross section which works together with our one meter vortex core mass. The maximum and minimum possible values of $\sigma/m$ are given as
\begin{align}
\rm{max}:\;\;\sigma/m &\approx 3.00\times10^{-18}\;\rm{cm^2}/\rm{g},\\
\rm{min}:\;\;\sigma/m &\approx 5.14\times10^{-19}\;\rm{cm^2}/\rm{g}.
\end{align}
The value of $\sigma/m$ for a thermalization radius of $84.7\,\rm{kpc}$, as presented in equation (\ref{eq:sigm_r1}), lies within these boundaries.

As a final remark, we should note that in imposing the constraint in equation (\ref{eq:m_constraint_ul}), we have also revisited the initial constraint of section \ref{sec:mass_test}.

%-------------------------------- The Temperature ------------------------------------
\subsubsection{The Temperature}
We have found values for which the vortices in the center of the superfluid DM have a core size of $\sim 1\,\rm{m}$, the halo density profile looks as expected, and the model constraints seem to be fulfilled. We now want to check if the halo temperature still is below the critical temperature for the new parameter values. For simplicity, we will again assume that the density is constant and equal to $\rho_{200}$, and take the same approach as in Section \ref{sec:temp}. The results are presented in Table \ref{table:1} and as before we obtain a dark matter temperature lower than the critical temperature, consistent with the formation of a BEC.
\begin{table}[h!]
\caption{Critical temperature and halo temperature when using a DM particle mass as presented in equation (\ref{eq:4.3.1}), $R=203\,\rm{kpc}$ and $M=10^{12}M_{\odot}$.}             % title of Table
\label{table:1}      % is used to refer this table in the text
\centering                          % used for centering table
\begin{tabular}{|c |c| c|}        % centered columns (4 columns)
\hline%\hline                 % inserts double horizontal lines
    DM particle mass & $T_c$ & $T_{DM}$ \\    % table heading 
\hline                        % inserts single horizontal line
   $m \simeq 1.1\times10^{-4}\,\rm{eV}$ &    $0.34\,\rm{K}$  &  $0.12\,\mu\rm{K}$   \\
\hline                                   %inserts single line
\end{tabular}
\end{table}

%-------------------------------- The Critical Angular Velocity ------------------------------------
\subsubsection{The Critical Angular Velocity}
From Section \ref{sec:properties} we have that the angular velocity of a Milky Way-like galaxy is of the order $\sim10^{-16}\,\rm{s}^{-1}$. We also calculated the critical angular velocity using equation (\ref{eq:omega_c}) for the constant density halo. Looking at Figure \ref{fig:dens_profile}, the density profile is not that far from constant, and we will therefore again use equation (\ref{eq:omega_c}) to obtain an approximation of the magnitude of the critical angular velocity. This results in
\begin{align}
\Omega_c &\sim 10^{-36}\;\rm{s}^{-1},
\end{align}
which shows that the galaxy still rotates fast enough for vortex formation.

%-------------------------------- The Energy of a Single Vortex ------------------------------------
\subsubsection{The Energy of a Single Vortex}
\label{sec:energy_vort}
The motivation behind making the vortex cores bigger was that it required a smaller mass, which in turn should increase the energy per vortex, as calculated at the end of Section \ref{sec:properties}. The energy expression, given as
\begin{align}
E = \frac{4\pi n(0,0)}{3}\frac{\hbar^2}{m}Z\ln\Big(0.671\frac{R}{\xi_0}\Big),
\end{align}
is based on a single vortex at the center and requires the number density and coherence length at the center. Again, this expression originates from the HO consideration, which means that it is valid for a constant density DM halo. Around a single vortex core the density may be assumed constant, as the length scales are way smaller than the length scales over which the condensate density changes. In addition, this expression is based on a cylindrical condensate shape and not a spherical one. For a central vortex, we may imagine, as a simple approach, taking a cylindrical sample around the vortex core and from that calculating the energy of the vortex. From this, we may insert $Z=2R$, $n(0,0) = \rho(0)/m$, the particle mass and the central vortex core size. We then get 
\begin{align}
E &\approx 4.2\times10^{13}\;\rm{J},
\end{align}
which converted to a mass is
\begin{align}
m_{\rm{vortex}}&\approx 4.8\times10^{-4}\;\rm{kg}.
\end{align}
From this we see that a smaller particle DM mass, gives larger vortices and more ``mass'' per vortex. If we assume that we still have $\sim10^{26}$ vortices in the condensate, as we had for the constant density case, and add all the masses together, ignoring the fact that the size of the vortices range from $1-24\,\rm{m}$ from center to edge, we obtain a total mass of $\sim10^{22}\,\rm{kg}$. This is of the same order of magnitude as the Moon.

%-------------------------------- The alpha parameter ------------------------------------
\subsubsection{The $\alpha$ Parameter}
\label{sec:alpha}
In the superfluid DM model presented by \cite{original}, it is the superfluid phonons that hold the key to the MONDian behavior, in the sense that they are governed by the MOND action. The MONDian force, essential for assuring the correct behavior of the model at galaxy scales, is ensured by the way the phonons couple to the baryons. The interaction term is given in their equation (26) as
\begin{align}
\mathscr{L}_{\rm{int}} = -\alpha\frac{\Lambda}{M_{\rm{Pl}}}\theta\rho_b, \label{eq:5.3.3.1}
\end{align}
where $\rho_b$ is the baryon mass density, $\theta$ is the phonon scalar field, $M_{\rm{Pl}}$ is the Planck mass, $\Lambda$ is an energy scale originating from the MOND scalar action, and finally, $\alpha$ is a dimensionless parameter. 

In addition to the two previous parameters, $m$ and $\Lambda$, there is an extra parameter of the model - the coupling constant, $\alpha$. In section 4 of \cite{original}, the superfluid phonons and their relation to the baryons are studied further, which results in a relation between the $\alpha$ and $\Lambda$ parameters, necessary to reproduce the MOND critical acceleration, $a_0\simeq1.2\times10^{-8}\,\rm{cm}/\rm{s}^2$ (equation (2) in \cite{original}). The relation is given as follows: 
\begin{align}
\alpha^{3/2}\Lambda = \sqrt{a_0M_{\rm{Pl}}} \simeq 0.8\;\rm{meV},\label{eq:5.3.3.2}
\end{align}
resulting in 
\begin{align}
\alpha \simeq 0.86\Big(\frac{\Lambda}{\rm{meV}}\Big)^{-2/3}.\label{eq:5.3.3.3}
\end{align}

From the relation in equation (\ref{eq:5.3.3.3}), it is clear that the smaller masses explored above will lead to very small values of $\alpha$. Based on equation (\ref{eq:5.3.3.1}), this is not something that invalidates our parameter variations, as it is the combination of $\alpha$ and $\Lambda$ that is of importance in the interaction term. 

%-------------------------------- Rotation Curves ------------------------------------
\subsubsection{Rotation Curves}
\begin{figure}
   \centering
   \includegraphics[trim={0 0 0 1.5cm},clip,scale=0.5]{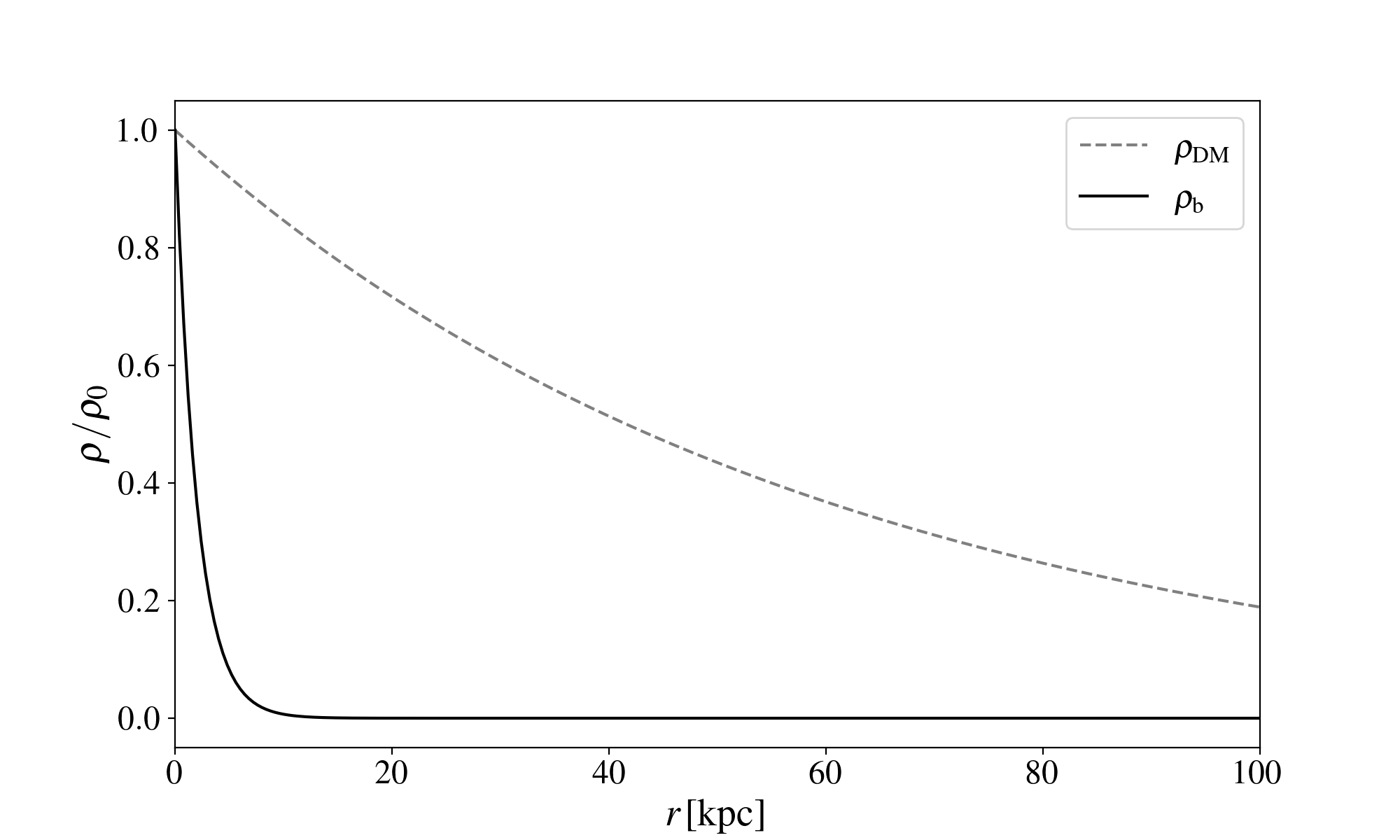}
      \caption{Toy density profiles used to produce mock rotation curves. Both the dark matter and the baryons follow the same distribution, only with a different spatial extent. The curves are normalized.
              }
         \label{fig:rot_c_densities}
   \end{figure}

A large part of the paper by \cite{Berezhiani2017} is devoted to showing that the SFDM model reproduces realistic rotation curves. Having changed the parameters, we must make sure that this is still the case. As our goal is not to reproduce the results of the aforementioned paper, but rather to figure out whether our new combinations of $m$ and $\Lambda$ will provide similar results as the original values, we may simplify the approach. Instead of solving for the dark matter density profile and fitting it to a Navarro-Frenk-White (NFW) profile at the outskirts, as done in \cite{Berezhiani2017}, we will choose a simplified spherical density profile (a toy profile), along with using a toy spherical density profile for the baryons. In \cite{Berezhiani2017}, equation (41), a toy profile for baryons is presented:
\begin{align}
\rho_b^{toy}(r) = \frac{M_b}{8\pi L^3}e^{-r/L},\label{eq:5.4.1}
\end{align}
where $L=2\,\rm{kpc}$ is a radial length scale defining how far out in the galaxy we want most of the mass to be distributed. We will adopt this profile for the baryons, and also for the dark matter,
\begin{align}
\rho_{DM}^{toy}(r) = \frac{M_{DM}}{8\pi L_{DM}^3}e^{-r/L_{DM}},\label{eq:5.4.2}
\end{align}
although with a different length scale, $L_{DM} = 60\,\rm{kpc}$. This is chosen to make sure that the dark matter is distributed far enough out to produce a flat rotation curve. The profiles are illustrated in Figure \ref{fig:rot_c_densities}.

\begin{figure}
   \centering
   \includegraphics[trim={0 0 0 1.5cm},clip,scale=0.5]{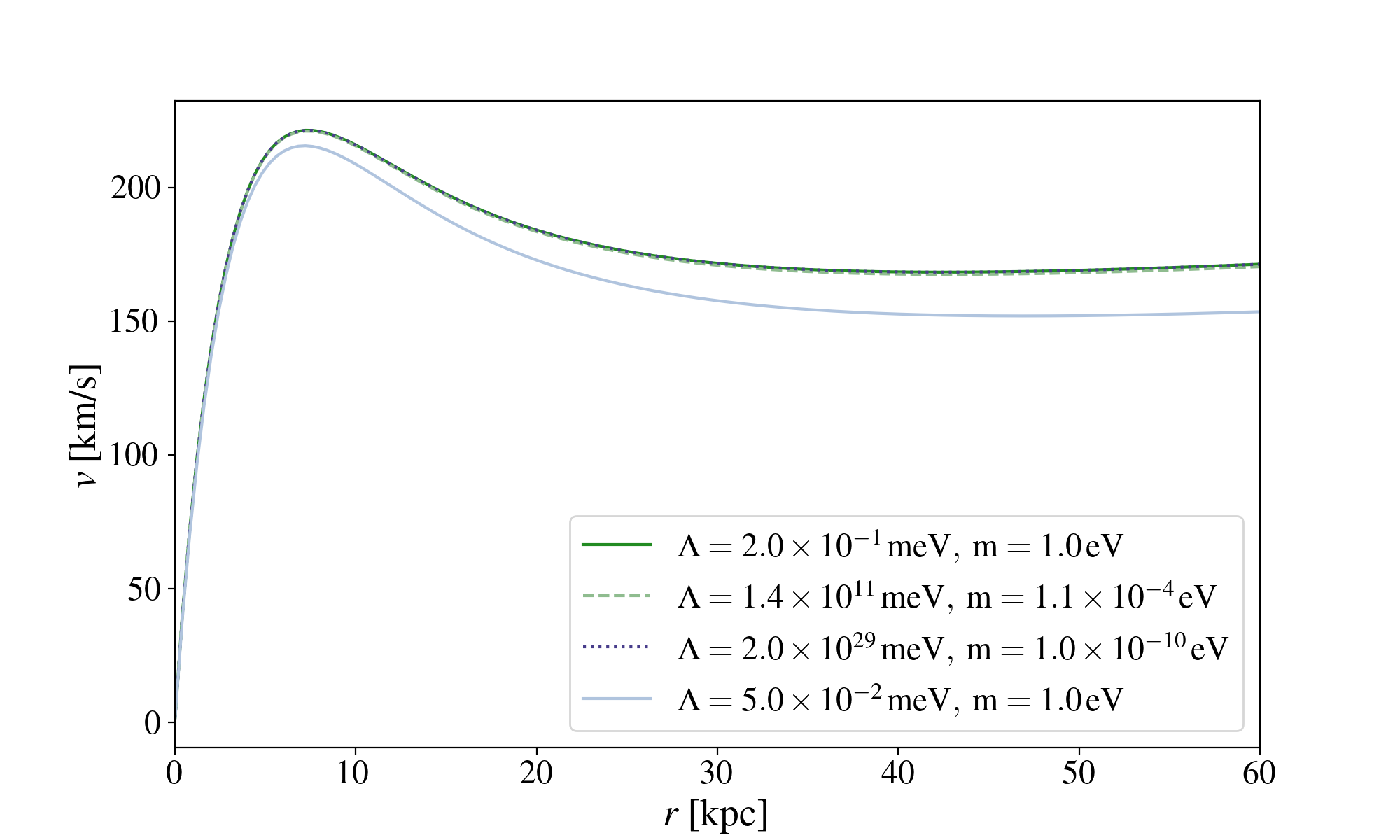}
      \caption{Resulting rotation curves of the toy density profiles presented in equation (\ref{eq:5.4.1}) and (\ref{eq:5.4.2}) for various combinations of $m$ and $\Lambda$. The green and purple lines result from the original $m$ and $\Lambda$ values, the one meter vortex core combinations and a combination that will be explored later in section \ref{sec:mass_var}. The light blue line is a result of the parameters $m = 1\,\rm{eV}$ and $\Lambda = 0.05\,\rm{meV}$, which have not been decided based on the same process as the others, and do not fulfill equation (\ref{eq:lambda_constraint}).
              }
         \label{fig:rot_c_vel}
   \end{figure}

The radial acceleration on a test (baryonic) particle inside the galaxy is given as
\begin{align}
a = a_b(r) + a_{DM}(r) + a_{\rm{phonon}}(r),\label{eq:5.4.3}
\end{align}
where
\begin{align}
a_b(r) &= GM_b(r)/r^2\nonumber\\
a_{DM}(r) &= GM_{DM}(r)/r^2\label{eq:5.4.4}\\
a_{\rm{phonon}}(r) &= \alpha\frac{\Lambda}{M_{\rm{Pl}}}\frac{d\theta}{dr},\nonumber
\end{align}
and $\theta$ is the phonon scalar field, governed by equation (32) in \cite{Berezhiani2017}:
\begin{align}
\frac{(\vec{\nabla}\theta)^2 + 2m\big(\frac{2\beta}{3}-1\big)\hat{\mu}}{\sqrt{(\vec{\nabla}\theta)^2 + 2m(\beta-1)\hat{\mu}}}\vec{\nabla}\theta = \alpha M_{\rm{Pl}}\vec{a}_b\label{eq:5.4.5}.
\end{align}
Here $\beta$ is a dimensionless constant introduced to parametrise finite-temperature effects\footnote{The condensate profile calculated in section \ref{sec:profile} was for the zero temperature case, but in reality the dark matter will have some finite small temperature.}, set to $\beta=2$ in the aforementioned paper. We also have that $\hat{\mu} = \mu - m\Phi$, where $\mu$ is the chemical potential and $\Phi$ is the gravitational potential. The chemical potential will be calculated based on equation (48) of \cite{original}
\begin{align}
\mu = \frac{\rho_{\rm{center}}^2}{8\Lambda^2m^5}\label{eq:5.4.6},
\end{align}
where $\rho_{\rm{center}}$ is given by equation (\ref{eq:central_density}) as before. This is based on the zero temperature case, but as our density profiles already are estimates we will still use it. The gravitational potential will be calculated as
\begin{align}
\nabla^2\Phi = 4\pi G(\rho_{DM} + \rho_b),\label{eq:5.4.7}
\end{align}
assuming rotational symmetry. Equation (\ref{eq:5.4.5}) may be rewritten as a cubic equation of $(\vec{\nabla}\theta)^2$, 
\begin{align}
0 = &(\vec{\nabla}\theta)^6 + 4m\Big(\frac{2\beta}{3}-1\Big)\hat{\mu}(\vec{\nabla}\theta)^4 + 4m^2\Big(\frac{2\beta}{3}-1\Big)^2\hat{\mu}^2(\vec{\nabla}\theta)^2 \nonumber\\
& -\alpha^2M_{\rm{Pl}}^2\vec{a}^2_b(\vec{\nabla}\theta)^2 - 2m(\beta-1)\hat{\mu}\alpha^2M_{\rm{Pl}}^2\vec{a}^2_b,\label{eq:daym}
\end{align}
where roots may be found numerically. For $m$ and $\Lambda$-values that fulfill the constraints, only one root will be real. To solve equation (\ref{eq:5.4.3}), we need the total mass of the galaxy along with an initial value of the gravitational potential at the center. The total baryonic mass is set to $M_b = 10^{11}M_{\odot}$, and the total dark matter mass to $10M_b$. The initial gravitational potential value is in \cite{Berezhiani2017} decided so that we obtain the total mass when integrating the density over the full volume of the galaxy. We choose the initial value in the same way. Once equation (\ref{eq:5.4.3}) has been solved, we may use the formula
\begin{align}
a = \frac{v_{\rm{circ}}^2}{r}\label{eq:5.4.9},
\end{align}
to obtain the corresponding rotation curve. Figure \ref{fig:rot_c_vel} shows the rotation curves for our original parameters, $m = 1\,\rm{eV}$ and $\Lambda = 0.2\,\rm{meV}$, along with the one meter center parameters, and one smaller mass that will be investigated in section \ref{sec:mass_var}. The rotation curve with the values $m = 1\,\rm{eV}$ and $\Lambda = 0.05\,\rm{meV}$ is also included.
The variations in parameters result in the same rotation curves. This indicates that an alteration in the parameters of the model should be possible and does not remove the models ability to reproduce realistic rotation curves. This should, however, also be checked for a more accurate dark matter density profile and actual baryonic density distribution observations. The last curve (light blue) is included to illustrate that we get a different rotation curve when we change $\Lambda$ without changing the mass.

%-------------------------------- DM Particle Mass Variation ------------------------------------
\subsubsection{DM Particle Mass Variation}
\label{sec:mass_var}
In Section \ref{sec:mass_test} and \ref{sec:energy_vort} we saw that a smaller DM particle mass leads to vortices that are both larger in spatial extent and energy content. As we are interested in vortices energetic enough to affect the host galaxy, a different but equivalent approach to examining the parameter space resulting in larger vortex cores would be to look at a range of smaller DM particle masses. This is what will be investigated in this section, with the specific DM particle masses $m=10^{-7}\,\rm{eV}$, $m=10^{-10}\,\rm{eV}$ and $m=10^{-15}\,\rm{eV}$. In \cite{original}, the DM candidate which enables the superfluidity is described as ``axion-like'' with a mass in the eV range. The mass of the axion has been proposed to be as low as $\sim10^{-22}\,\rm{eV}$ in fuzzy dark matter models \citep{hu2000}, but is more conventionally thought to be in the $\sim10^{-5}-10^{-3}\,\rm{eV}$ range \citep{klaer}. %In addition, we now know that smaller particle masses, and by that larger vortex cores, result in a higher vortex energy. This again gives each vortex a higher ``mass,'' which further increases their gravitational pull, making any interaction between the vortices and baryons more likely.

\begin{figure}
\centering
\subfloat[\normalsize$m=10^{-7}\,\rm{eV}$]{\centerline{\includegraphics[trim={0 0 0 1.1cm},clip,scale=0.47]{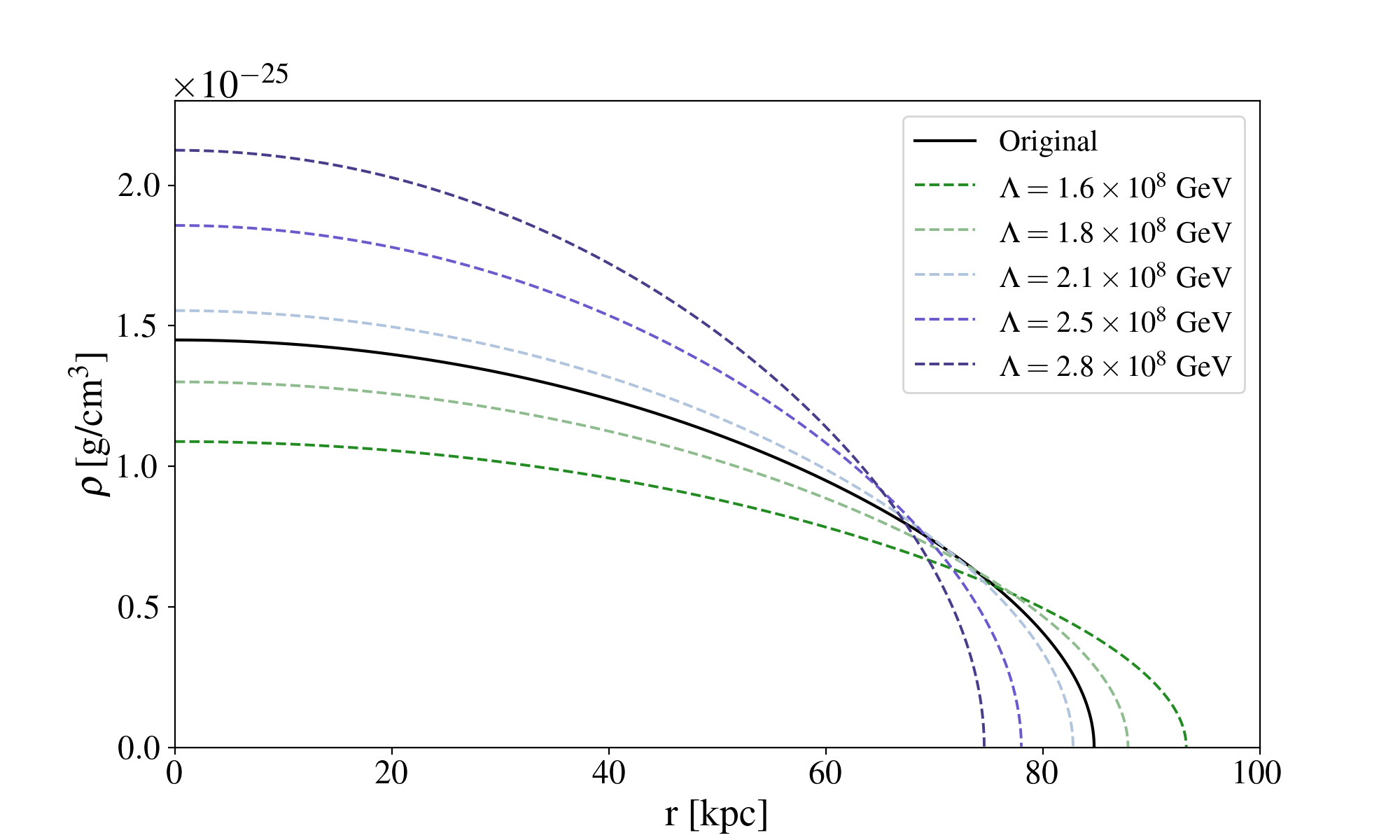}}}\\[-0.3ex]
\subfloat[\normalsize$m=10^{-10}\,\rm{eV}$]{\centerline{\includegraphics[trim={0 0 0 1.1cm},clip,scale=0.47]{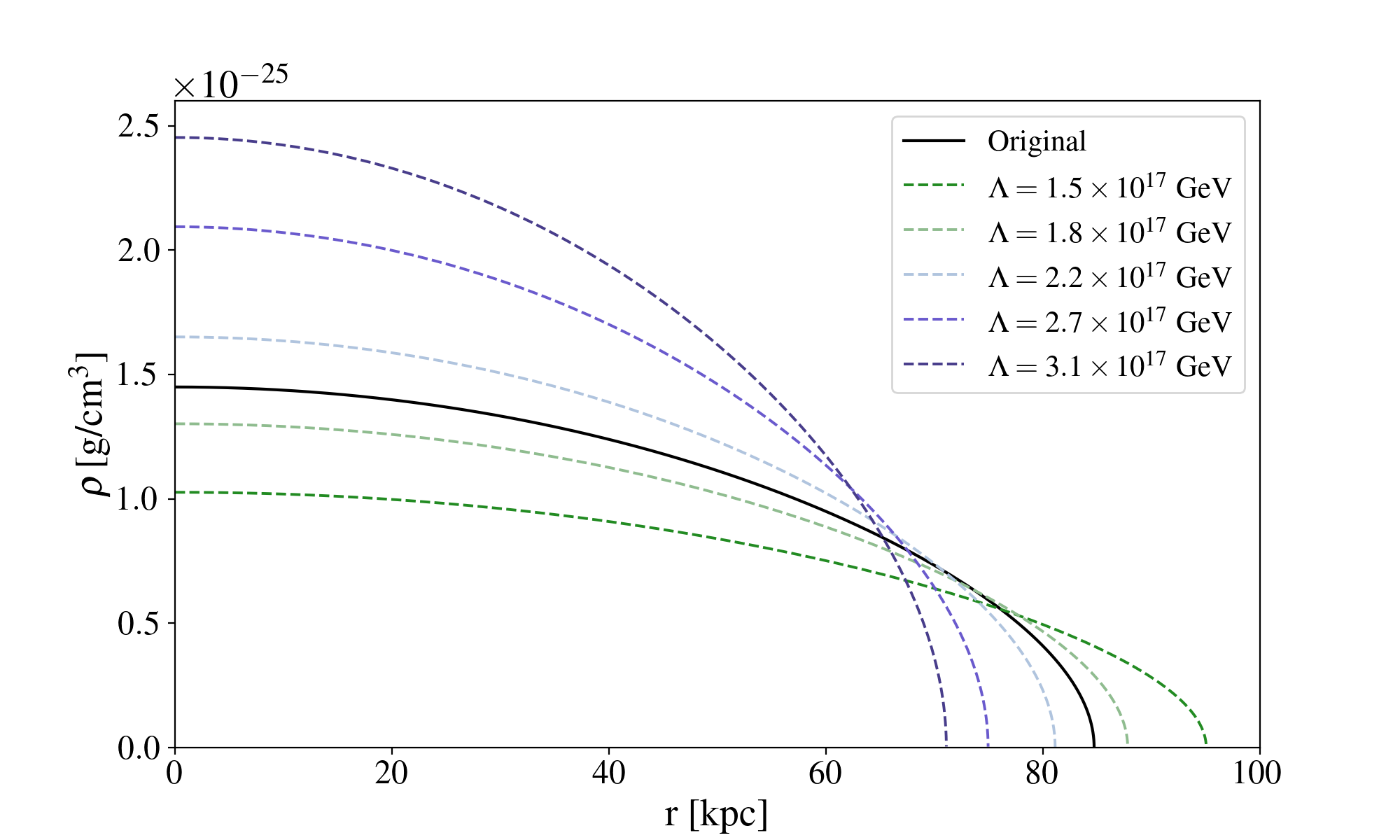}}}\\[-0.3ex]
%\end{figure}
%\begin{figure}[H]
%\ContinuedFloat
%\centering
\subfloat[\normalsize$m=10^{-15}\,\rm{eV}$]{\centerline{\includegraphics[trim={0 0 0 1.1cm},clip,scale=0.47]{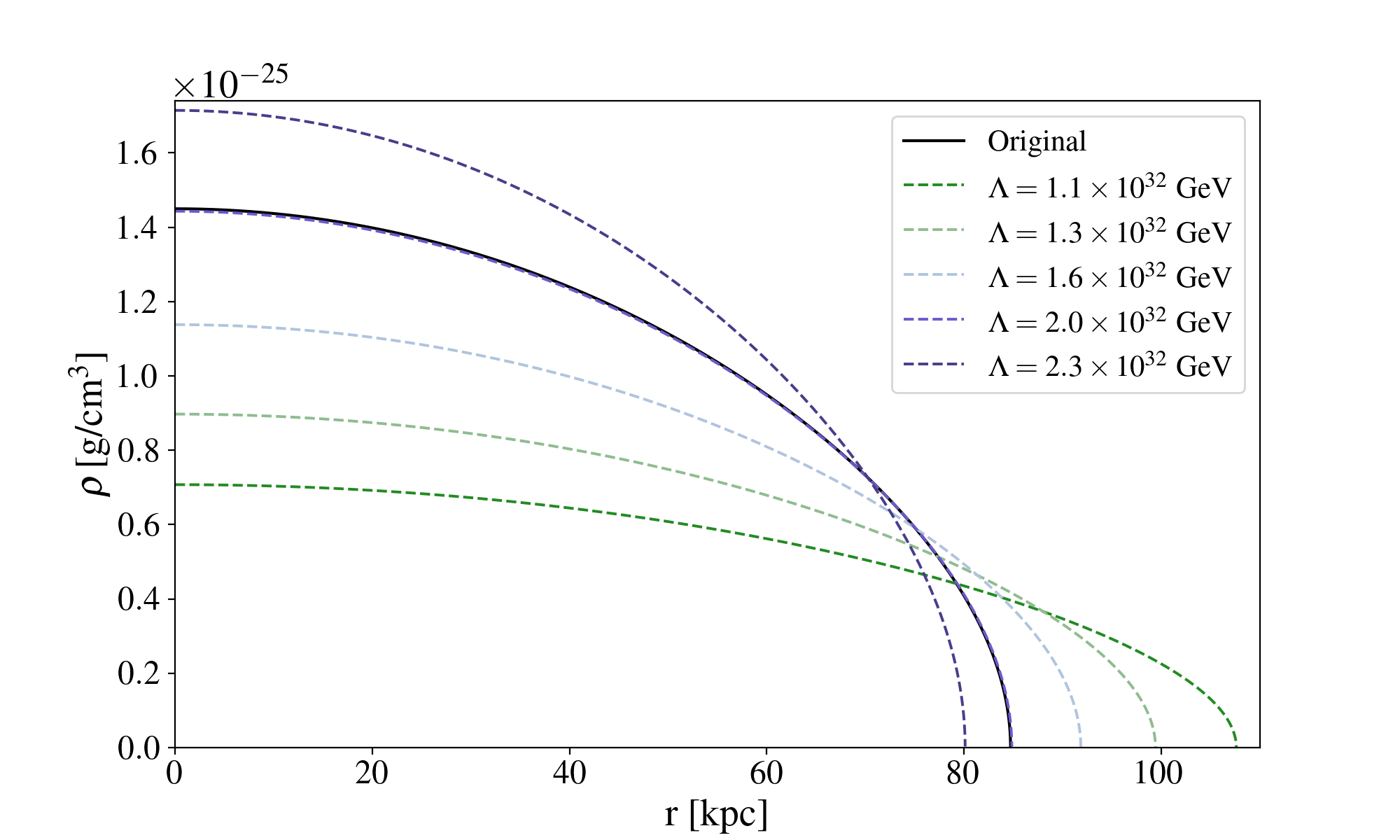}}}
%\caption[]{-- Continued.}
\caption[Dark Matter Density Profile Fitting for Various Values of $m$.]{Dark matter density profile fitting for various values of $m$. The lower the value of the DM particle, the higher the value of $\Lambda$ required to reproduce the original density profile. It is worth noting that the mass of $m=10^{-15}\,\rm{eV}$ requires a $\Lambda$-value well beyond the reduced Planck mass, $M_{\rm{Pl}}\approx2.435\times10^{18}\,\rm{GeV}$.}
\label{fig:many_m_values}
\end{figure}

In Figure \ref{fig:many_m_values}, we have gone through the same process of choosing a value of $m$ and finding a value of $\Lambda$ that fits the density profile, as was done in section \ref{sec:lambda_value}. For each of the new mass values we also calculate the vortex core size and the energy/mass of a single vortex at the center of the condensate. An overview of the $\Lambda$-values required by the different masses is given in Table \ref{tab:lambda_table}, along with the vortex core size, energy, mass, intervortex spacing and the number of vortices. We previously calculated the two latter variables for our simple constant density halo, but as the intervortex spacing depends on the DM particle mass as shown in equation (\ref{eq:intervortex}) and the number of vortices again depends on the intervortex spacing, we reevaluate these values for each mass. The table also includes the DM particle mass resulting in a one meter vortex core, for easier comparison. 
%\begin{table*}
%\caption{Overview of model parameters and resulting vortex properties.}             
%\label{tab:lambda_table}      
%\centering          
%\begin{tabular}{|c| c| c| c| c| c| c|}     % 7 columns 
%\hline%\hline       
                      % To combine 4 columns into a single one 
%DM particle mass  &    $\Lambda$-value  & Vortex core & Vortex energy &   Vortex ``mass''& $a_{\Omega}$ & N$_{\rm{vortex}}$\\ 
%\hline                 
 %  $1.1\times10^{-4}\,\;\rm{eV}$  & $\approx141\,\;\rm{MeV}$  & $1.0\,\;\rm{m}$&$4.2\times10^{13}\,\rm{J}$ &$4.8\times10^{-4}\;\,\rm{kg}$&$1.7\times10^{-1}\,\rm{AU}$& $\sim10^{22}$\\
 
 %  $10^{-7}\,\;\rm{eV}$  & $\sim2\times10^{8}\,\;\rm{GeV}$  & $1.1\times10^3\,\;\rm{m}$&$4.7\times10^{19}\,\rm{J}$ &$5.2\times10^2\;\,\rm{kg}$& $5.8\times10^{0}\,\rm{AU}$& $\sim10^{19}$\\
 
 %  $10^{-10}\,\rm{eV}$   &    $\sim2\times10^{17}\,\rm{GeV}$  & $1.1\times10^6\,\;\rm{m}$&$4.7\times10^{25}\,\rm{J}$ &$5.2\times10^8\;\,\rm{kg}$ & $1.8\times10^{2}\,\rm{AU}$& $\sim10^{16}$\\
 
 %  $10^{-15}\,\rm{eV}$   &    $\sim2\times10^{32}\,\rm{GeV}$  & $1.1\times10^{11}\,\rm{m}$&$4.7\times10^{35}\,\rm{J}$ &$5.2\times10^{18}\,\rm{kg}$ & $5.8\times10^{4}\,\rm{AU}$& $\sim10^{11}$\\
%\hline                  
%\end{tabular}
%\end{table*}

\begin{table*}
\caption{Overview of model parameters and resulting vortex properties.}             
\label{tab:lambda_table}      
\centering          
\begin{tabular}{|c| c| c| c| c|}     % 7 columns 
\hline%\hline       
                      % To combine 4 columns into a single one 
DM particle mass  &   $1.1\times10^{-4}\,\;\rm{eV}$ & $10^{-7}\,\;\rm{eV}$ & $10^{-10}\,\rm{eV}$ & $10^{-15}\,\rm{eV}$\\\hline
$\Lambda$-value  & $\approx141\,\;\rm{MeV}$ & $\sim2\times10^{8}\,\;\rm{GeV}$ & $\sim2\times10^{17}\,\rm{GeV}$ & $\sim2\times10^{32}\,\rm{GeV}$\\
Vortex core & $1.0\times10^0\,\;\rm{m}$ & $1.1\times10^3\,\;\rm{m}$ & $1.1\times10^6\,\;\rm{m}$ & $1.1\times10^{11}\,\;\rm{m}$\\
Vortex energy & $4.2\times10^{13}\,\rm{J}$ &  $4.7\times10^{19}\,\rm{J}$ & $4.7\times10^{25}\,\rm{J}$ & $4.7\times10^{35}\,\rm{J}$ \\
Vortex ``mass''& $4.8\times10^{-4}\;\,\rm{kg}$ & $5.2\times10^2\;\,\rm{kg}$ & $5.2\times10^8\;\,\rm{kg}$ & $5.2\times10^{18}\;\,\rm{kg}$\\
$a_{\Omega}$ &$1.7\times10^{-1}\,\rm{AU}$ & $5.8\times10^{0}\,\rm{AU}$ & $1.8\times10^{2}\,\rm{AU}$ & $5.8\times10^{4}\,\rm{AU}$ \\
N$_{\rm{vortex}}$ & $\sim10^{22}$ & $\sim10^{19}$ & $\sim10^{16}$ & $\sim10^{11}$\\

%\hline                 
 %  $1.1\times10^{-4}\,\;\rm{eV}$  & $\approx141\,\;\rm{MeV}$  & $1.0\,\;\rm{m}$&$4.2\times10^{13}\,\rm{J}$ &$4.8\times10^{-4}\;\,\rm{kg}$&$1.7\times10^{-1}\,\rm{AU}$& $\sim10^{22}$\\
  % $10^{-7}\,\;\rm{eV}$  & $\sim2\times10^{8}\,\;\rm{GeV}$  & $1.1\times10^3\,\;\rm{m}$&$4.7\times10^{19}\,\rm{J}$ &$5.2\times10^2\;\,\rm{kg}$& $5.8\times10^{0}\,\rm{AU}$& $\sim10^{19}$\\
  % $10^{-10}\,\rm{eV}$   &    $\sim2\times10^{17}\,\rm{GeV}$  & $1.1\times10^6\,\;\rm{m}$&$4.7\times10^{25}\,\rm{J}$ &$5.2\times10^8\;\,\rm{kg}$ & $1.8\times10^{2}\,\rm{AU}$& $\sim10^{16}$\\
  % $10^{-15}\,\rm{eV}$   &    $\sim2\times10^{32}\,\rm{GeV}$  & $1.1\times10^{11}\,\rm{m}$&$4.7\times10^{35}\,\rm{J}$ &$5.2\times10^{18}\,\rm{kg}$ & $5.8\times10^{4}\,\rm{AU}$& $\sim10^{11}$\\
\hline                  
\end{tabular}
\end{table*}

From the table we see that the resulting $\Lambda$-value for the $m=10^{-10}\,\rm{eV}$ case gets close to the reduced Planck mass of $M_{\rm{Pl}}=\sqrt{\hbar c/8\pi G}\approx2.435\times10^{18}\,\rm{GeV}$. For $m=10^{-15}\,\rm{eV}$, the $\Lambda$-value needed to fit the original density profile is immense, $10^{14}$ times the Planck mass scale. The Planck mass is in itself no upper or lower limit, but reaching these energy levels usually indicates that the theory is not complete or may not be well understood without the use of quantum gravity. Based on this, there is no need to go even further down to $m\sim10^{-22}\,\rm{eV}$.

\begin{figure}
   \centering
   \includegraphics[trim={0 0 0 1.5cm},clip,scale=0.5]{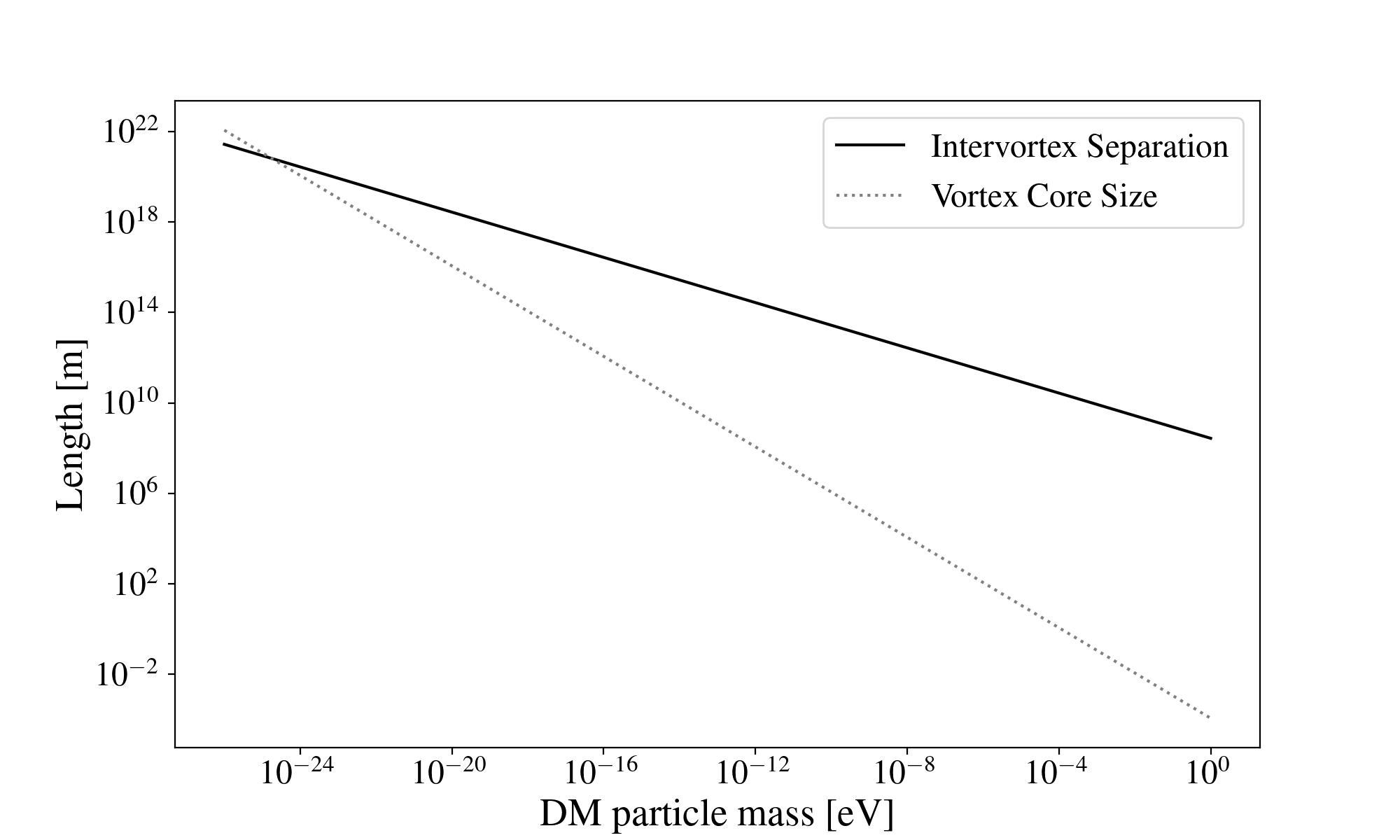}
      \caption{DM particle mass dependency of the intervortex separation and vortex core size (the axis are logarithmic). The two values overlap at $m\approx1.8\times10^{-25}\;\rm{eV}$.
              }
         \label{fig:overlap}
   \end{figure}

Table \ref{tab:lambda_table} also shows that we may obtain very large vortex cores and corresponding vortex ``masses'' by lowering the DM particle mass. In addition we notice that the intervortex separation increases with decreasing DM particle mass, further resulting in fewer vortices in the condensate as a whole. As both the vortex core size and intervortex separation grows as the DM particle mass is reduced, one might wonder when they will overlap. 
%However, for the smallest DM particle mass, we reach a vortex core size larger than the separation of the vortices in the condensate, if we assume that the vortex separation does not change drastically from the constant density case. This means that the vortices would overlap and that we would need a new description of the system. As a comparison, 
For type II superconductors, the superconductive state would cease to exist when the vortex cores start overlapping. For a Bose-Einstein condensate, however, vortex lattices can remain stable even in these situations (see section 9.5 of \cite{pethick_smith_2008}). Figure \ref{fig:overlap} shows how the intervortex separation and vortex core size depends on the DM particle mass. 
The two lines, in logarithmic space, intersects at a DM particle mass of $m\approx1.8\times10^{-25}\;\rm{eV}$. As we have already seen that $m=10^{-15}\,\rm{eV}$ gives a $\Lambda$-value way above the Planck mass scale, there is no reason to investigate the DM particle mass at which the vortex cores starts overlapping in this first approach. 

%Still, combined with the fact that the $\Lambda$-value is way above the Planck mass scale, a DM particle mass as low as $10^{-15}\,\rm{eV}$ seems improbable in the current SFDM model. 
The $10^{-10}\,\rm{eV}$ particle mass is also on the border of Planck mass scales, and yet all $10^{16}$ vortices combined only amount to a mass of $\sim 1M_{\oplus}$. The Milky Way contains $\sim10^{11}$ stars, and the gravitational impact of an extra Earth-like planet is negligible. Adding all the vortices of the $m=10^{-7}\,\rm{eV}$ case together, we end up at a total mass of $\sim 10^{21}\;\rm{kg}$, which is roughly the mass of Charon. 

At this point, we must remember that the expressions used to calculate the vortex core size and energy were based on a constant halo density. Previously, we have continued using the formulas as the vortices have been so small that the density could be assumed constant in their proximity. At least for the lowest DM particle mass, we can no longer assume the density to be constant, and the numbers presented in Table \ref{tab:lambda_table} must therefore be treated with caution.

It should be noted that there exists other models describing a Bose-Einstein condensate DM, one of which has vortices on kpc scale \cite{zinner}. In the given paper, however, the condensate is dominated by two-body interactions, as opposed to the SFDM model where three-body interactions must dominate to obtain the necessary MONDian force. In addition, the DM particle mass considered in the aforementioned paper is $m \sim 10^{-22}\;\rm{eV}$, which for us would lead to $\Lambda$-values way above the Planck mass scale. Still, if we calculate the vortex core size for a DM particle mass of $m \sim 10^{-22}\;\rm{eV}$ we obtain a value of $\sim 0.03\;\rm{kpc}$ in the center of the condensate. We would have to consider even lower DM particle masses to get a vortex core of a similar size as in \cite{zinner}. If we go all the way down to the mass where the vortex cores would start overlapping ($m\approx1.8\times10^{-25}\;\rm{eV}$), we get a core size of $\sim 20\;\rm{kpc}$. At this point, however, we would have to find a new description of our system. 

From these simple tests, it is clear that it is possible to find a $\Lambda$-value that will reproduce the desired density profile for very small values of $m$. However, how small the particle mass can get is limited by the fact that $\Lambda$ increases significantly as the mass is decreased, quickly leading us beyond Planck energy scales. We have also seen that too low values of $m$ will lead to the vortex cores overlapping, which would change the description of the system, and possibly affect the phonons needed to mediate the MONDian force. 

%-------------------------------- Superfluid Dark Matter Lagrangian ------------------------------------
\section{Superfluid Dark Matter Lagrangian}
\label{sec:SFDML}
In \cite{original} Berezhiani and Khoury suggest a relativistic completion of the SFDM model, 
\begin{align}
\mathscr{L} = -\frac{1}{2}(|\partial_{\mu}\phi|^2 + m^2|\phi|^2) - \frac{\Lambda^4}{6(\Lambda_c^2 + |\phi|^2)^6}(|\partial_{\mu}\phi|^2 + m^2|\phi|^2)^3.\label{eq:7.1}
\end{align}
The parameters are as before, and $\Lambda_c$ is introduced to allow for a vacuum solution ($\phi=0$). Equation (\ref{eq:7.1}) is a relativistic completion of equation (\ref{eq:modelll}), in the zero temperature limit. In this section we will explore a possible vortex solution of the given Lagrangian.

%-------------------------------- Equations of Motion ------------------------------------
\subsection{Equations of Motion}
Our goal is to extract a vortex equation from the full relativistic SFDM Lagrangian. To do so we will first express the condensate wavefunction, $\phi$, in terms of a modulus, $\rho(X)$, and phase, $\psi(X)$,
\begin{align}
\phi = \frac{\rho(X)}{\sqrt{2}}e^{i\psi(X)}\label{eq:7.1.2},
\end{align}
as done in \cite{schmitt2014}. Here $X=(t,r,\theta,z)$ denotes spacetime polar coordinates, using natural units. The phase is described as $\psi(X) = n\theta + \mu t$, where $n$ is an integer and $\mu$ is the chemical potential. Moving forward we will use the metric signature (-,+,+,+) and assume a static solution, $\partial_0\rho=0$. Writing out the Lagrangian we now have
\begin{align}
\mathscr{L} = &-\frac{1}{4}\Big(g^{\mu\nu}\partial_{\mu}\rho\partial_{\nu}\rho + g^{\mu\nu}\rho^2\partial_{\mu}\psi\partial_{\nu}\psi + m^2\rho^2\Big) \notag\\ 
&- \frac{\Lambda^4}{48(\Lambda_c^2+\frac{1}{2}\rho^2)^6}\Big(g^{\mu\nu}\partial_{\mu}\rho\partial_{\nu}\rho + g^{\mu\nu}\rho^2\partial_{\mu}\psi\partial_{\nu}\psi + m^2\rho^2\Big)^3.\label{eq:7.1.4}
\end{align}
In \cite{schmitt2014}, a vortex solution is found by studying the Euler-Lagrange equation with respect to $\rho$:
\begin{align}
\frac{\partial\mathscr{L}}{\partial\rho} - \partial_{\eta}\frac{\partial\mathscr{L}}{\partial(\partial_{\eta}\rho)} = 0 \label{eq:7.1.9}.
\end{align}
The same approach will be taken here. We will assume $\rho(X)=\rho(r)$ and use $g_{\mu\nu}=\rm{diag}(-1,1,r^2,1)$. We also define the parameter
\begin{align}
q = m^2-\mu^2+\frac{n^2}{r^2}\label{eq:7.1.14}
\end{align}
and use the notation ``$\;\dot{ }\;$'' to represent the derivative with respect to $r$. The resulting equation of motion is 
\begingroup
%\allowdisplaybreaks
\setlength{\abovedisplayskip}{17pt}
\setlength{\belowdisplayskip}{17pt}
\begin{align}
0=& -\frac{1}{2}q\rho +\frac{\Lambda^4\rho}{8(\Lambda_c^2 + \frac{1}{2}\rho^2)^7}\big(\dot{\rho}^2 + q\rho^2\big)^3 
-\frac{\Lambda^4}{8(\Lambda_c^2 + \frac{1}{2}\rho^2)^6}q\rho\big(\dot{\rho}^2 + q\rho^2\big)^2\notag\\
&+\frac{1}{2}\ddot{\rho} - \frac{3\Lambda^4\rho\dot{\rho}^2}{4(\Lambda_c^2 + \frac{1}{2}\rho^2)^7}\big(\dot{\rho}^2 + q\rho^2\big)^2 + \frac{\Lambda^4}{8(\Lambda_c^2 + \frac{1}{2}\rho^2)^6}\ddot{\rho}\big(\dot{\rho}^2 + q\rho^2\big)^2\label{eq:7.1.16}\\
&+\frac{\Lambda^4\dot{\rho}}{4(\Lambda_c^2 + \frac{1}{2}\rho^2)^6}\Big[2\dot{\rho}\ddot{\rho} -2\frac{n^2}{r^3}\rho^2 + 2q\rho\dot{\rho}\Big]\big(\dot{\rho}^2 + q\rho^2\big)\notag.
\end{align}
\endgroup
The next step is to write equation (\ref{eq:7.1.16}) on unitless form. To do so we introduce the variables 
\begin{align}
R=\rho/\rho_0 \quad\rm{and}\quad \eta=\rho_0r.\label{eq:new_var}
\end{align}
Here we will let $\rho_0$ be the solution to the Euler-Lagrange equations with respect to $\rho$ when $\nabla\rho=\nabla\psi=0$, for the case $\Lambda_c\ll |\phi|^2$ (the MOND regime). This procedure is inspired by \cite{schmitt2014} and results in 
\begin{align}
\rho_0 &= \sqrt{2\Lambda}(m^2-\mu^2)^{1/4}\label{eq:7.1.27}.
\end{align}
To calculate the chemical potential we use equation (48) in \cite{original}, 
\begin{align}
\mu = \frac{\rho_{\rm{center}}^2}{8\Lambda^2m^5},
\end{align}
where $\rho_{\rm{center}}$ denotes the central density of the condensate (equation \ref{eq:central_density}). Inserting the new variables into equation (\ref{eq:7.1.16}) and rearranging gives
\begingroup
\allowdisplaybreaks
\setlength{\abovedisplayskip}{20pt}
\setlength{\belowdisplayskip}{10pt}
\begin{align}
\ddot{R} = \frac{f(\eta, R, \dot{R})}{g(\eta, R, \dot{R})}\label{eq:7.1.28},
\end{align}
\endgroup
with
\begingroup
%\allowdisplaybreaks
\setlength{\abovedisplayskip}{20pt}
\setlength{\belowdisplayskip}{20pt}
\begin{align}
f(\eta, R, \dot{R}) =\:& \rho_0^2(5R\dot{R}^6 + 9pR^3\dot{R}^4 + 3p^2R^5\dot{R}^2 -p^3R^7)\notag\\
&-\Big(\Lambda_c^2 + \frac{1}{2}\rho_0^2R^2\Big)\Big[-\frac{4p}{\Lambda^4\rho_0^8}R\Big(\Lambda_c^2 + \frac{1}{2}\rho_0^2R^2\Big)^6 \label{eq:7.1.29}\\
&+ 3pR\dot{R}^4 + 2p^2R^3\dot{R}^2 - p^3R^5 - 4R^2\dot{R}\frac{n^2}{\eta^3}(\dot{R}^2 + pR^2)\Big],\notag\\
\quad\notag\\
g(\eta, R, \dot{R}) =\;& \Big(\Lambda_c^2 + \frac{1}{2}\rho_0^2R^2\Big)\Big[\frac{4}{\Lambda^4\rho_0^8}\Big(\Lambda_c^2 + \frac{1}{2}\rho_0^2R^2\Big)^6 \notag\\&+ 5\dot{R}^4 + 6pR^2\dot{R}^2 + p^2R^4\Big]\label{eq:7.1.30}.
\end{align}
\endgroup
Here we have defined 
\begingroup
\allowdisplaybreaks
\setlength{\abovedisplayskip}{5pt}
\setlength{\belowdisplayskip}{20pt}
\begin{align}
p &= \frac{\rho_0^2}{4\Lambda^2}+\frac{n^2}{\eta^2}\label{eq:7.1.31},
\end{align}
\endgroup
and the ''$\;\dot{ }\;$`` represents the derivative with respect to $\eta$. 

%-------------------------------- Interpretation ------------------------------------
\subsection{Interpretation}
Equation (\ref{eq:7.1.28}) is a second-order nonlinear ordinary differential equation with associated boundary conditions. Any solution similar to a known vortex solution should have the boundary conditions $R(\eta=0)=0$ and $R(\eta\to\infty)\to1$. A full solution of equation (\ref{eq:7.1.28}) is yet to be found, and this section is therefore devoted to investigating the behavior of $R(\eta)$ in the limit of small and large values of $\eta$, or equivalently close and far away from the supposed vortex core. For this we will use the values $m = 1\,\rm{eV}$, $\Lambda = 0.2\,\rm{meV}$ and $\Lambda_c = \Lambda/2$. Based on the results, we will also present an analysis of how different initial values and $\Lambda_c$-values impact the solution of the equation.

%-------------------------------- Small eta ------------------------------------
\subsubsection{Small $\eta$}
\begin{figure}
\centering
\subfloat[]{\includegraphics[trim={0 0 1.5cm 1cm},clip,scale=0.4]{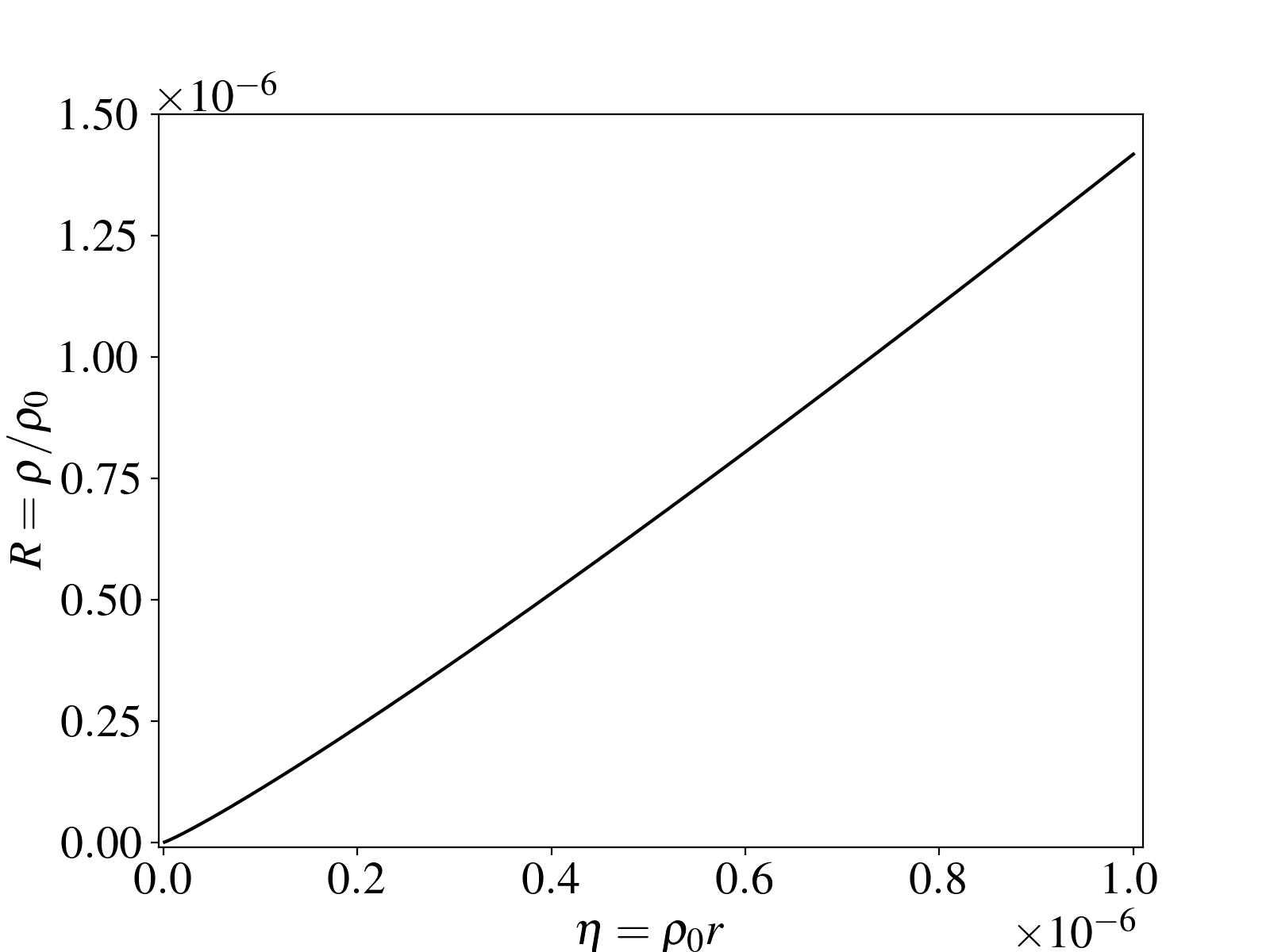}}
\subfloat[]{\includegraphics[trim={0 0 0 1cm},clip,scale=0.4]{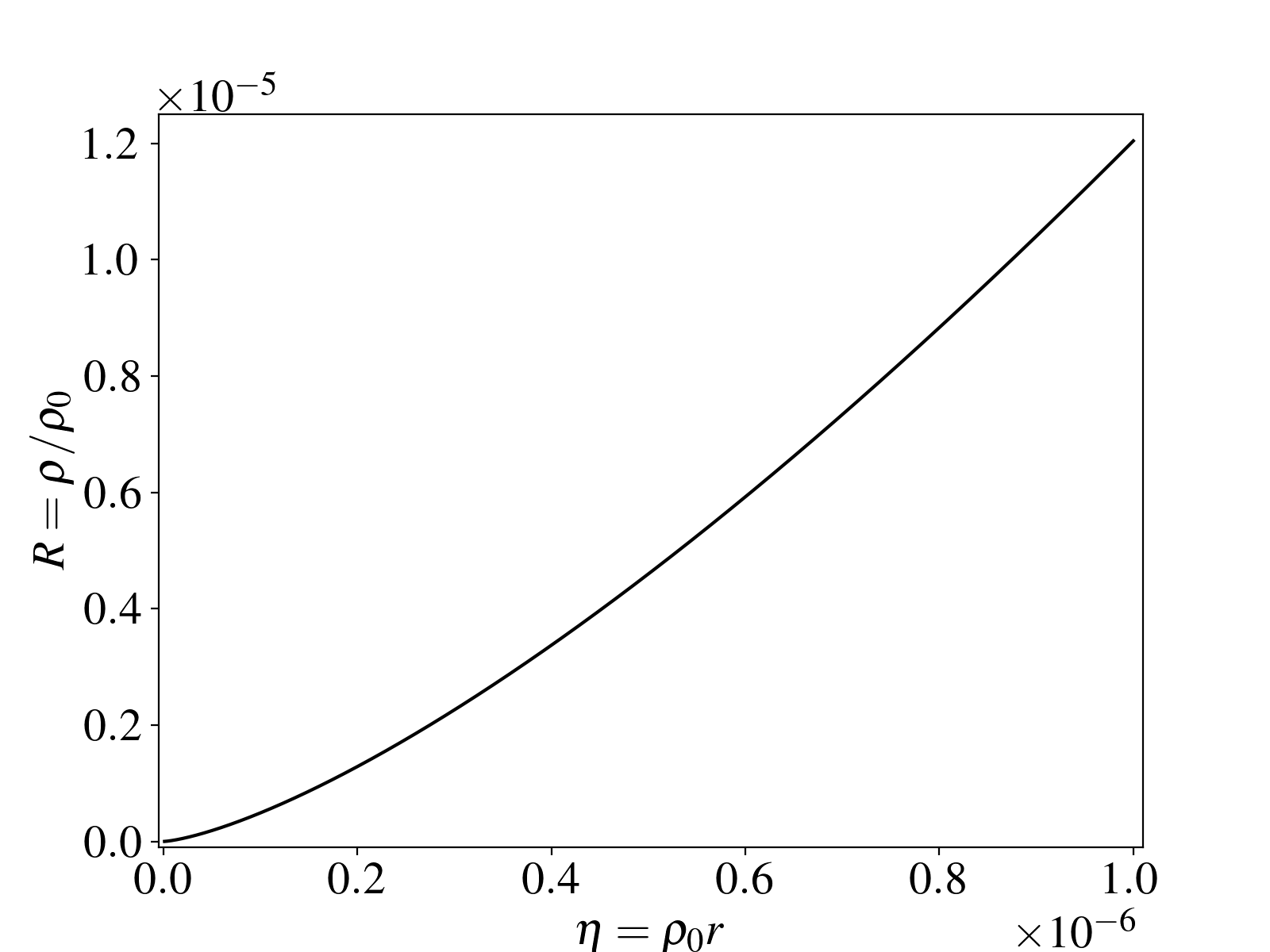}}
\caption{Figure (a) shows the solution of equation (\ref{eq:7.1.28}) for very small values of $\eta$, using the approximate vortex solution ($R = \eta/\sqrt{2+\eta^2}$) to decide the initial conditions. Figure (b) again shows the solution close to the core, but now using $R=\eta$ and $\dot{R}=1$ as initial conditions.}
\label{fig:small_eta}
\end{figure}

First we look at the behavior of equation (\ref{eq:7.1.28}) for small values of $\eta$. If there is a vortex solution, this would be close to the vortex core. To figure out which terms dominate equation (\ref{eq:7.1.28}) in this limit, we must make some assumptions about $R$ and $\dot{R}$. The simplest approach is to assume the existence of a vortex solution and use the approximate vortex profile $R(\eta)=\eta/\sqrt{2+\eta^2}$ from \cite{pethick_smith_2008} to estimate initial values. Using a value $\eta=10^{-9}$, corresponding to $r\sim10^{-13}\,\rm{cm}$, we find the following approximate expression for small $\eta$:
\begin{align}
\ddot{R} &\sim \frac{-\Big(\Lambda_c^2 + \frac{1}{2}\rho_0^2R^2\Big)\Big(3pR\dot{R}^4 - 4R^2\dot{R}\frac{n^2}{\eta^3}\dot{R}^2\Big)}{\Big(\Lambda_c^2 + \frac{1}{2}\rho_0^2R^2\Big)5\dot{R}^4}\label{eq:7.3.2.1}.
\end{align}
Simplifying and only including the second term in $p$ ($p\sim n^2/\eta^2$), along with inserting $n=1$, we get 
\begin{align}
\ddot{R} &\sim \frac{R}{\eta^2}\Big(\frac{3}{5} - \frac{4}{5}\frac{R}{\eta\dot{R}}\Big)\label{eq:7.3.2.2}.
\end{align}
Equation (\ref{eq:7.3.2.2}) is solved numerically using a 4th order Runge-Kutta method. The solution is shown in figure \ref{fig:small_eta}a, and it displays an almost linear behavior, which is in accordance with a regular vortex solution close to the core.

It could be the case that there is no vortex solution or that the solution is on an unfamiliar form. We will therefore also take a look at the small $\eta$ limit without using the approximate vortex solution as an initial condition. We will assume a linear relation, $\eta=R$, in the given limit, along with $\dot{R}\gg R$. This gives the approximate expression
\begin{align}
\ddot{R} &\sim \frac{4R^2\dot{R}^2\eta^2 + 4R^4}{5\dot{R}^3\eta^5+6R^2\dot{R}\eta^3}\label{eq:7.3.2.3}
\end{align}
when inserting $n=1$ and $p\sim n^2/\eta^2$ as before. Again we use $\eta=10^{-9}$ as our starting point, and we also choose $\dot{R}=1$. The result is shown in figure \ref{fig:small_eta}b. The new profile is slightly different, but very similar to the previous one. As any increasing function would look linear if we go to small enough x-values, the behavior at large $\eta$ will be the most interesting in this case.

%-------------------------------- Large eta ------------------------------------
\subsubsection{Large $\eta$}
\label{sec:large}
In general one would expect a vortex profile to converge towards a constant value far from the center. In this limit we are in the regime $\Lambda_c \ll |\phi|$, which in \cite{original} is presented as the MOND regime of the theory. To analyze this limit we may first put $\Lambda_c=0$ in equation (\ref{eq:7.1.28}). Then we again assume that a vortex solution similar to $R(\eta)=\eta/\sqrt{2+\eta^2}$ exists, and use the given expression to estimate the values of the different terms of equation (\ref{eq:7.1.28}) far from the core. From earlier we have that the radius of the vortex core in the center of the condensate is $\sim10^{-4}\,\rm{m}$, which corresponds to $\eta\approx10$. We therefore choose a value $\eta=200$ to analyze the equation in the far away limit. Only keeping the dominant terms along with letting $p\to\frac{\rho_0^2}{4\Lambda^2}$ results in the equation of motion
\begin{align}
\ddot{R} \sim \frac{\rho_0^2}{4\Lambda^2}R\label{eq:7.3.3.1}
\end{align}
in the large $\eta$ limit. Immediately we can see that this corresponds to an exponential profile, and therefore does not represent a possible vortex solution. %The result is illustrated in figure \ref{fig:large_eta}a \textcolor{red}{REMOVE FIGURE?}.
%\begin{figure}
%\centering
%\subfloat[]{\includegraphics[scale=0.39]{large_eta_vort_02_new.pdf}}\\
%\subfloat[]{\includegraphics[scale=0.39]{large_eta_below1_02_new.pdf}}
%\caption{Figure (a) shows the solution of equation (\ref{eq:7.1.28}) for large values of $\eta$, using the approximate vortex solution ($R = \eta/\sqrt{2+\eta^2}$) to decide the initial conditions. Figure (b) again shows the solution far from the core, but now using $R=0.97$ and $\dot{R}=10^{-6}$ as initial conditions. \textcolor{red}{Was $\Lambda_c$ set to zero in the calculations of these solutions? Does not seem like it from the code. The exponential solution goes away then.. but how? Based on eq 85 it shouldnt...}}
%\label{fig:large_eta}
%\end{figure}

\begin{figure}
\centering
\includegraphics[trim={0 0 0 1.5cm},clip,scale=0.5]{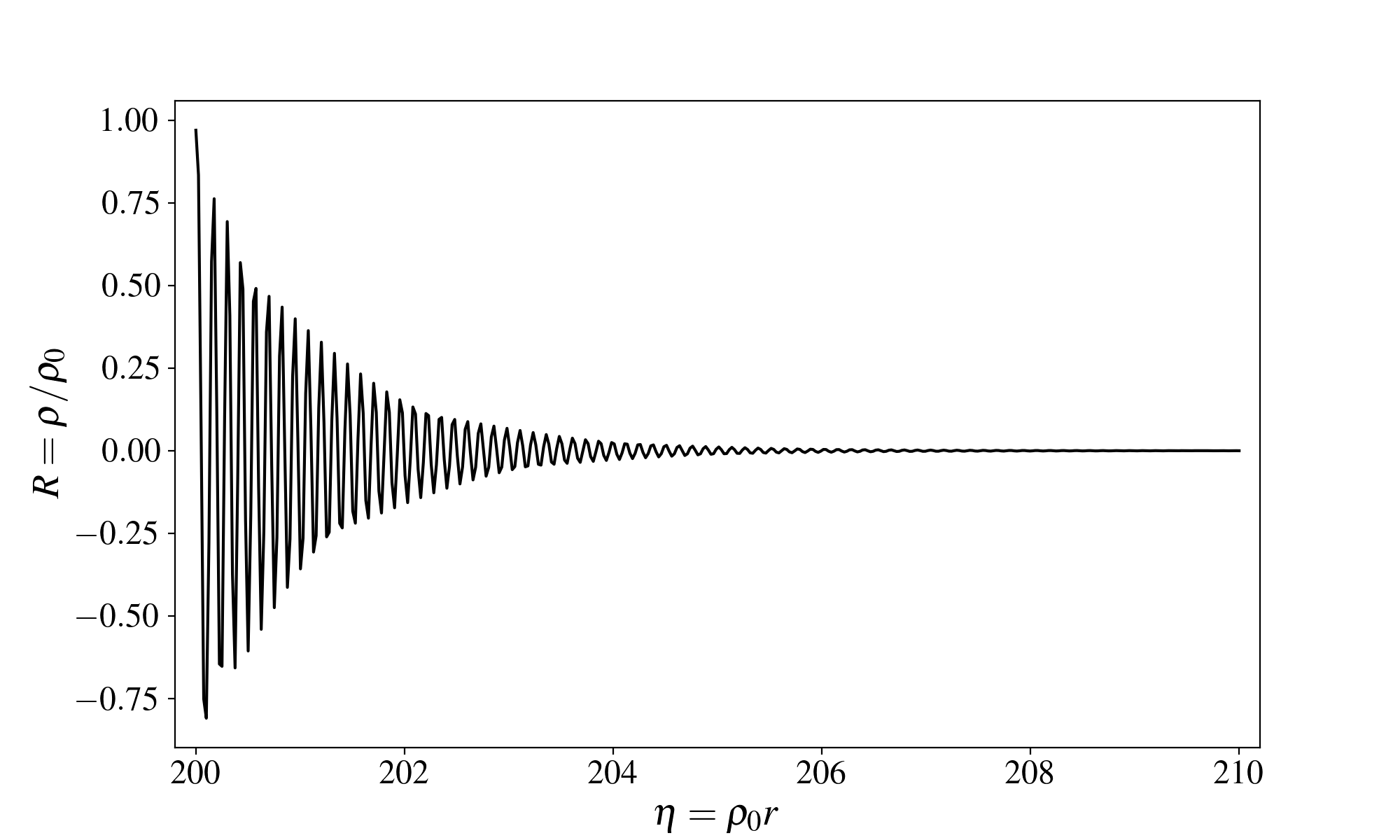}
\caption{Solution far from the core, using $R=0.97$ and $\dot{R}=10^{-6}$ as initial conditions. %\textcolor{red}{Was $\Lambda_c$ set to zero in the calculations of these solutions? Does not seem like it from the code. The exponential solution goes away then.. but how? Based on eq 85 it shouldnt...}}
}
\label{fig:large_eta}
\end{figure}

To obtain the expression in equation (\ref{eq:7.3.3.1}) we used the approximation $p\to\frac{\rho_0^2}{4\Lambda^2}$. If we instead use the full expression for $p$ and choose initial values of $R$ and $\dot{R}$ ourselves, we may obtain solutions on the form illustrated in Figure \ref{fig:large_eta}. This profile oscillates around zero, and settles down to a constant value. This solution is, as the other, not a possible vortex profile. There are, however, several examples where vortex profiles have oscillations close to the singularity \citep{galli, ancilotto}, although this is usually a property of very dense superfluids, such as He\Romannum{2} \citep{villerot}. The superfluid dark matter is a dilute superfluid and is not expected to have this property. Still, studying a superfluid in the dark matter picture could bring with it unexpected behavior.

In conclusion, we find that the vortex equation (\ref{eq:7.1.28}) is unstable at large $\eta$, and any vortex solution on a familiar form seems improbable. However, it is also clear that the form of the solution is very sensitive to the choice of initial conditions.

%-------------------------------- Numerical Stability ------------------------------------
\subsubsection{Numerical Stability}
\begin{figure}
   \centering
   \includegraphics[trim={0 1.5cm 0 1.5cm},clip,scale=0.6]{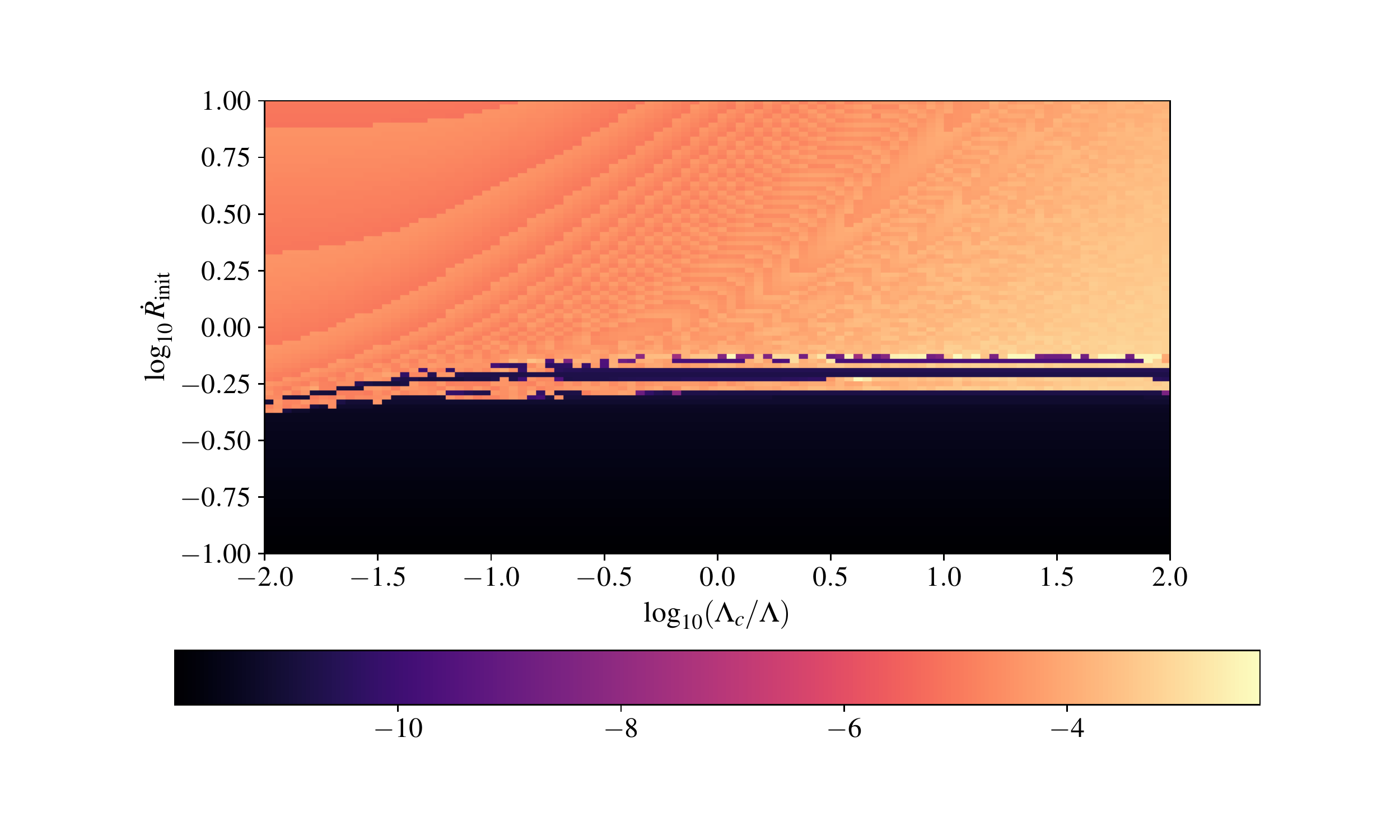}
      \caption{Grid showing how ``good'' a shot is, based on how many iterations are completed without the value of $R$ exceeding $R=2$. The equations have been normalized to $R\to1$. We allow higher values to make sure that solutions that overshoot in the beginning, but later stabilize, are included. How ``well'' a shot has done is indicated by a color gradient, moving from black to white. The scale is logarithmic, meaning that $0$ would indicate a solution that has converged to the given boundary values, while the size of the negative values indicate how far along the $\eta$-axis the completed iterations take us, before surpassing $R=2$ or diverging in some way or another. The solutions are weighted by the number $(i/N)\times R_{\rm{max}}$, where $i$ is the number of iterations reached, $N$ is the total number of iterations and $R_{\rm{max}}$ is the maximum value of the solution. A good shot would have $i\approx N$ and $R_{\rm{max}}\approx1$, so that the total weight is close to 1. It is the logarithm of this value that is displayed in the colorbar at the bottom.
              }
         \label{fig:grid}
   \end{figure}

If we look at the limit $\eta\to\infty$, and assume that the solution behaves as expected, we should have $\dot{R}\to0$ and $R\to1$. Inserting this into equation (\ref{eq:7.1.29}) and (\ref{eq:7.1.30}), along with $p\to\frac{\rho_0^2}{4\Lambda^2}$, and keeping \textit{all} terms, we get:
\begingroup
\allowdisplaybreaks
\setlength{\abovedisplayskip}{5pt}
\setlength{\belowdisplayskip}{20pt}
\begin{align}
f&\to -\rho_0^2\frac{\rho_0^6}{4^3\Lambda^6} - \Big(\Lambda_c^2 + \frac{1}{2}\rho_0^2\Big)\Big[-\frac{4\frac{\rho_0^2}{4\Lambda^2}}{\Lambda^4\rho_0^8}\Big(\Lambda_c^2 + \frac{1}{2}\rho_0^2\Big)^6 -\frac{\rho_0^6}{4^3\Lambda^6}\Big],\\
g&\to \Big(\Lambda_c^2 + \frac{1}{2}\rho_0^2\Big)\Big[\frac{4}{\Lambda^4\rho_0^8}\Big(\Lambda_c^2 + \frac{1}{2}\rho_0^2\Big)^6 + \frac{\rho_0^4}{4^2\Lambda^4}\Big].
\end{align}
\endgroup
In this limit we should also have $\Lambda_c\ll |\phi|$, which from earlier means that we can ignore $\Lambda_c$ in our expressions. This gives
\begin{align}
f\to& -\frac{\rho_0^8}{2^6\Lambda^6} -  \frac{1}{2}\rho_0^2\Big[-\frac{1}{\Lambda^6\rho_0^6}\frac{1}{2^6}\rho_0^{12} -\frac{\rho_0^6}{2^6\Lambda^6}\Big]=0,\\
g\to& \,\frac{1}{2}\rho_0^2\Big[\frac{4}{\Lambda^4\rho_0^8}\frac{1}{2^6}\rho_0^{12} + \frac{\rho_0^4}{4^2\Lambda^4}\Big]= \frac{\rho_0^6}{2^4\Lambda^4},
\end{align}
which is self-consistent and indicates that there could be a solution. Because of this, we have tried to solve equation (\ref{eq:7.1.28}) directly in various ways. One approach is the \textit{shooting method}. Here you guess at one of the initial conditions of your boundary value problem, solve the equation like a regular initial value problem and then check if the last value is far from a given boundary condition. If the value is within a given tolerance, your shot was successful. If not, you make a new guess and try again. Any vortex profile should fulfill the boundary conditions $\rho(r=0)=0$ and $\rho(r\to\infty)\to\rm{constant}$. We can therefore use the shooting method and check if the final solution has a constant value for large $\eta$, and if not, solve it again with new initial conditions. The problem with this is mentioned in Section \ref{sec:large}. Because of the possible exponential nature of the equation in the large $\eta$ limit, many initial values will immediately make the ODE-solver result in an overflow, thereby removing the possibility of checking if the last point is close to the desired boundary condition, and from that make a new guess at the initial values. We will therefore use the shooting method to test the stability of equation (\ref{eq:7.1.28}) and the parameter space of $\dot{R}_{\rm{init}}$ and $\Lambda_c$. This will be done by making several initial guesses at $\dot{R}_{\rm{init}}$ for various values of $\Lambda_c$ and checking if the solution fulfills the desired boundary values of a vortex profile. The results are illustrated in Figure \ref{fig:grid}.

\begin{figure}
   \centering
   \includegraphics[trim={0 1.5cm 0 1.2cm},clip,scale=0.6]{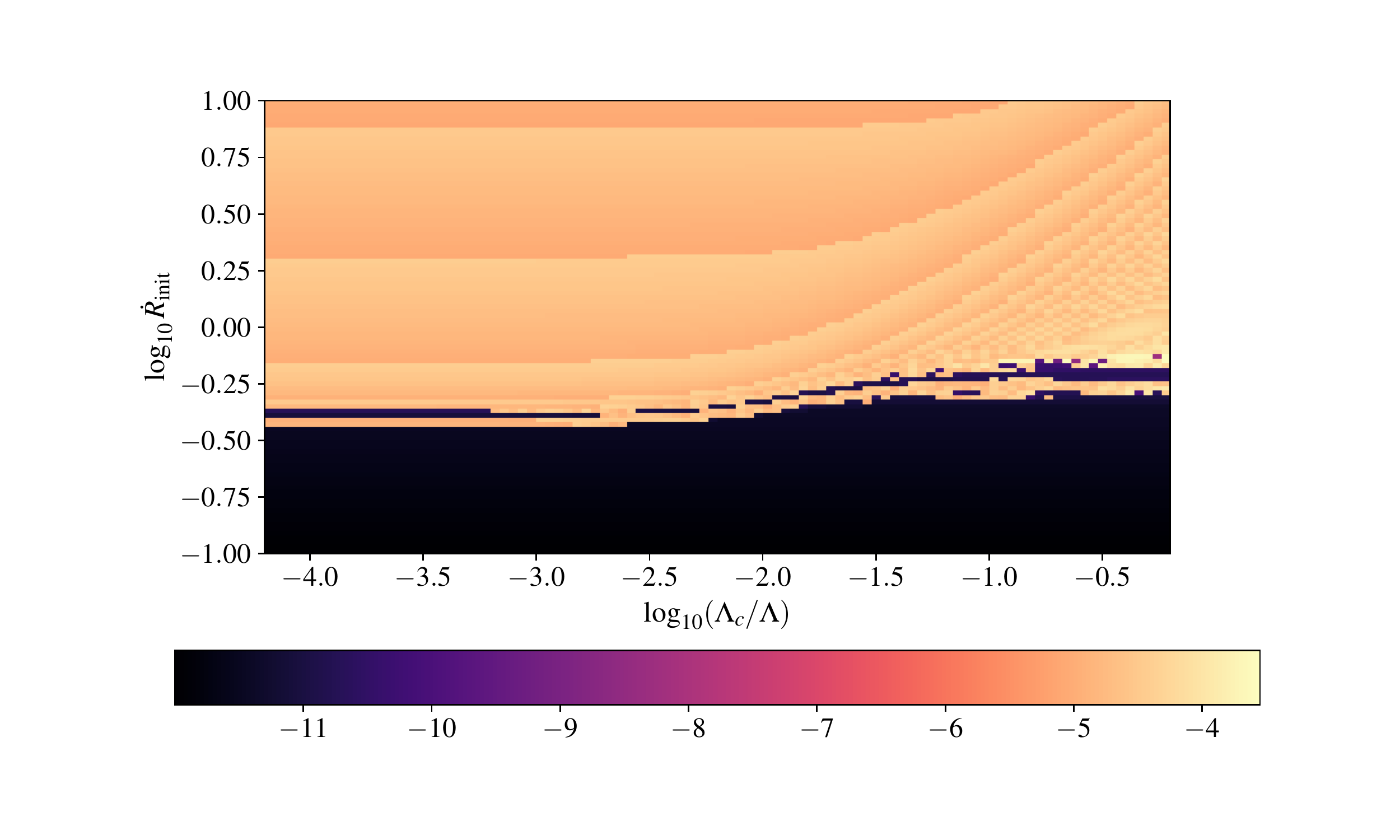}
      \caption{Remake of Figure \ref{fig:grid}, excluding values of $\Lambda_c$ that have been ruled out by observations. The most promising region of Figure \ref{fig:grid} is now gone, leaving us with a small area of plausible parameter combinations for $\log_{10}\dot{R}_{\rm{init}}\in[-0.3,0.0]$ and $\log_{10}(\Lambda_c/\Lambda)\in[-1.0,0.0]$.
              }
         \label{fig:grid_real}
   \end{figure}

From Figure \ref{fig:grid} it is evident that most values diverge. The scale moves from black to white, where the black points are parameter combinations that diverge from the desired boundary conditions right away, and the brighter points survive a higher number of iterations before diverging. A few yellow points may be observed in the $\log_{10}\dot{R}_{\rm{init}}\in[-0.3,0.0]$ region, for values of $\log_{10}(\Lambda_c/\Lambda)\in[0.0,2.0]$. These represent shootings that could provide vortex-like solutions. However, observational constraints on the MOND regime result in an upper limit of $\Lambda_c$, given as $\frac{\Lambda_c}{\Lambda}\lesssim \frac{\alpha^2}{10}$. This may be found in section 6 of \cite{original}. For our value of $\Lambda=0.2\,\rm{meV}$, we have a value $\alpha \approx 2.51$ from section \ref{sec:alpha}, which again results in the limit $\Lambda_c/\Lambda \lesssim 0.63$. This means that all the points above $\log_{10}(\Lambda_c/\Lambda)\approx-0.2$ in Figure \ref{fig:grid} are ruled out by observations. This region is also where we find most of our plausible solutions to the vortex equation. Figure \ref{fig:grid_real} illustrates the leftover area.

\begin{figure}
   \centering
   \includegraphics[trim={0 1.5cm 0 1.5cm},clip,scale=0.6]{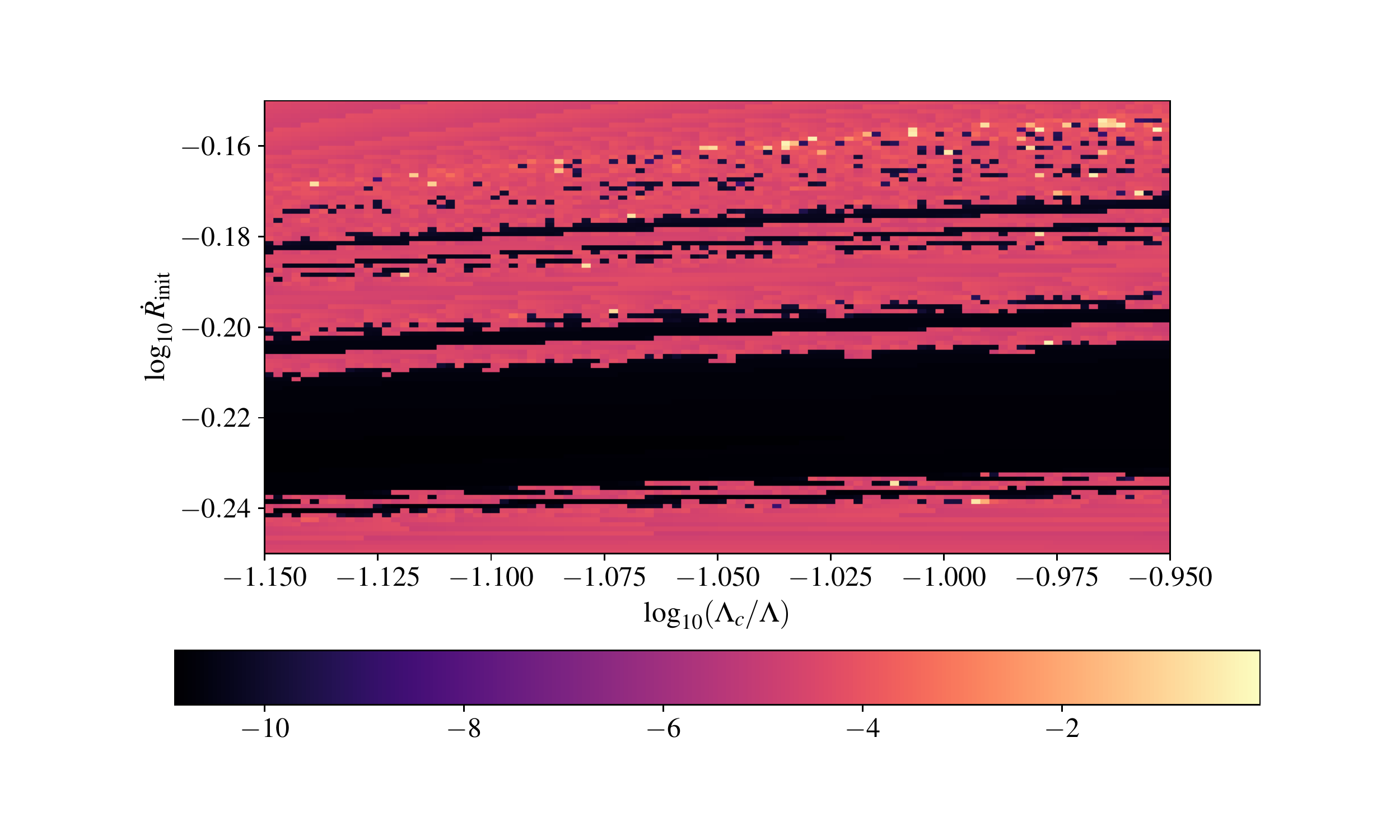}
      \caption{More refined grid for the region $\log_{10}\dot{R}_{\rm{init}}\in[-0.25,-0.15]$, $\log_{10}(\Lambda_c/\Lambda)\in[-1.15,-0.95]$ in Figure \ref{fig:grid_real}. It is evident that the given region is highly unstable, and that none of the shootings completely reach the desired results. There are only a few promising yellow points.}
         \label{fig:grid_zoom}
   \end{figure}

Investigating Figure \ref{fig:grid_real}, we find that the area $\log_{10}\dot{R}_{\rm{init}}\in[-0.25,-0.15]$ and $\log_{10}(\Lambda_c/\Lambda)\in[-1.15,-0.95]$ holds the most promising results. This is shown in Figure \ref{fig:grid_zoom}.
It is again clear that the given region is unstable, and that most of the attempts fail to give desirable results. There are a few light yellow points which reach values close to zero on the logarithmic scale. This means that the parameter combination still gives a diverging result, but that the given attempt survive a higher number of iterations. This could indicate that there exists, for some fine-tuned values of $\dot{R}_{\rm{init}}$ and $\Lambda_c$, a vortex solution with the desired boundary conditions.

\begin{figure}
   \centering
   \includegraphics[trim={0 0 0 1cm},clip,scale=0.5]{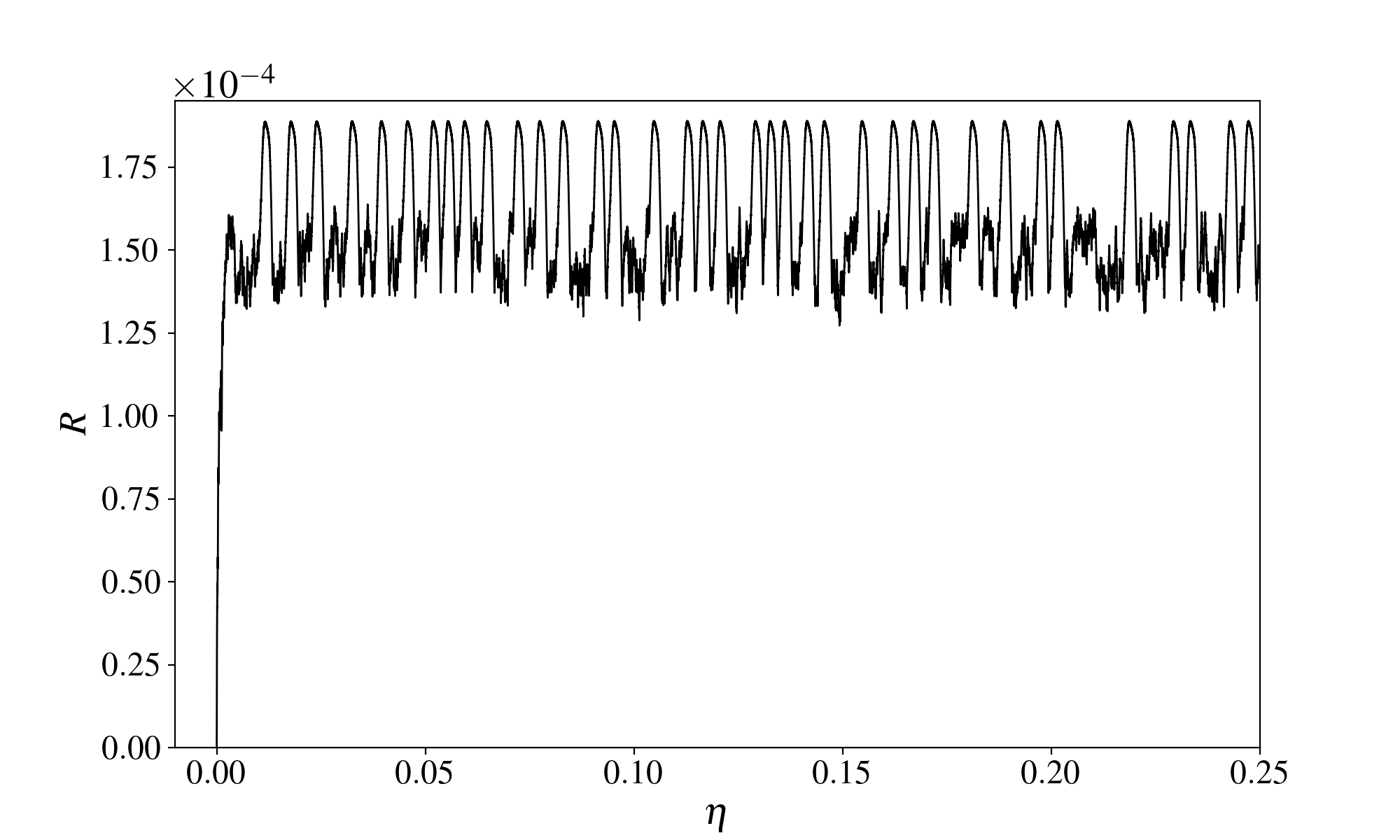}
      \caption{Example of a ``good'' shot from Figure \ref{fig:grid_zoom}. The plot shows the solution of the vortex equation for $\dot{R}_{\rm{init}}\simeq 0.693$ and $\Lambda_c\simeq0.018\,\rm{meV}$. The vortex profile has an oscillatory behavior, as also seen in section \ref{sec:large}, but will move on to diverge for $\eta\sim3$. In addition, the solution does not oscillate around $R=1$, but rather $R\sim1.5\times10^{-4}$.
              }
         \label{fig:grid_good}
   \end{figure}

In Figure \ref{fig:grid_good} an example of a ``good'' attempt is shown. 
It is clear that the resulting vortex profile is not a good solution. Firstly, the solution never reaches values close to $R=1$ and secondly, the profile diverges towards infinity at $\eta\sim3$. This is not included in the figure, in order to better illustrate the behavior for small $\eta$. This means that even though the shooting method provides ``good'' solutions, they do not necessarily serve as realistic vortex profiles. 

For now we have only investigated the parameter space of $\dot{R}_{\rm{init}}$ and $\Lambda_c$. As we have seen previously, the SFDM model allows for quite drastic changes in the parameters $m$ and $\Lambda$, which could possibly open up for other values of $\dot{R}_{\rm{init}}$ and $\Lambda_c$. Still, at this point, a realistic vortex solution of the theory presented in \cite{original} has not been found in this investigation. 

%---------------------------- Conclusions -------------------------
\section{Discussion and Conclusions}
In \cite{original} and \cite{Berezhiani2017} a new DM model is proposed, where the DM condenses and forms a superfluid on galactic scales. The superfluid exhibits collective excitations known as phonons, which interact with baryons and mediate a MONDian force. In this way, the SFDM model embodies a MONDian behavior on small scales, and a $\Lambda$CDM behavior on large scales. As a consequence of a rotating superfluid, we know from condensed matter physics that vortices should form in a grid throughout the condensate if the fluid rotates faster than some critical angular velocity. In this paper we have investigated vortex properties, explored the $m$ and $\Lambda$ parameter space of the SFDM model and attempted to find a vortex solution to the relativistic Lagrangian presented for the model in \cite{original}. In this section we will summarize and discuss our findings.

We started off our vortex investigation by approximating the SFDM halo to have a constant density. This meant that we could use already established knowledge about condensates trapped in a HO potential to estimate some vortex properties. From this we found:
\label{sec:conc}
   \begin{enumerate}
      \item The angular velocity of a Milky Way-like galaxy is above the critical angular velocity for vortex formation, and the DM temperature is below the critical temperature required for condensation to take place.
      \item The size of a single vortex core is approximately $1\,\rm{mm}$ and the vortex grid that should arise based on the angular velocity of a galaxy has an intervortex spacing of $\sim0.002\;\rm{AU}$.
      \item The condensate radius of our constant density halo is $\lesssim 83.16\,\rm{kpc}$.
      \item The energy per unit length required to produce a single vortex is $\sim 10^{-13}\,\rm{g}\,\rm{cm/s}^2$ and the corresponding ``mass'' of a vortex is approximately $10^{-14}\,\rm{kg}$. We remark in passing that this corresponds to a deficit angle of the order of $10^{-56}$ arcseconds if the vortex line is treated as a cosmic string \cite{peacock_1998}, so there will be no observable lensing effects. 
   \end{enumerate}
This analysis was based on a $200\rho_{\rm{crit}}$ constant density and a DM particle mass of $m=1\,\rm{eV}$. Based on the assumption that we have a uniform grid throughout the condensate, the size and spacing of two vortices can be compared to a flea on the Moon in relation to a flea on Earth. Combined with the fact that the ``mass'' per vortex is incredibly small and that all the $10^{26}$ vortices in this scenario add up to less than the mass of Mount Everest, it is unlikely that the vortices affect the surrounding baryons in any observable way. However, we have here assumed that the SFDM halo rotates uniformly, resulting in a uniform grid which mimics rigid body rotation. A galaxy does not rotate uniformly, and this could also be the case for the SFDM halo. If so, one could imagine that the vortex lattice would have a non-uniform configuration, which could have a different overall effect on the baryons in the galaxy. Ideally, one would like to simulate the full vortex grid within a galaxy to study the formation and effect of the vortices on the surroundings, but this is not possible due to the small size of the vortices in this scenario. 

%An experiment carried out in \cite{Sahu2018} shows that vortices formed in a condensate that does not rotate uniformly have a tendency to move towards the edge of the condensate in a ring formation. In reality, the density of the SFDM halo is not constant, which leads to larger and more ``massive'' vortices towards the edge of the condensate. It could therefore be the case that a a ring formation of vortices could have a larger overall effect on the baryons in the galaxy than a uniform grid. Ideally, one would like to simulate the full vortex grid within a galaxy to study the formation and effect of the vortices on the surroundings, but this is not possible due to the small size of the vortices in this scenario. 
   
Because of the seemingly unimportant vortices resulting from the SFDM model parameter choices in \cite{original} and \cite{Berezhiani2017}, we investigated the $m-\Lambda$ parameter space of the model, with the goal of obtaining more ``massive'' vortices which could have an observable impact on their surroundings. The idea is that the vortices might drag the surrounding baryons along and slightly affect their movements, which seems more likely if the vortices are large and have a greater gravitational impact. It is also interesting to investigate the parameter space in and of itself, as the axion particle mass varies greatly in different proposed theories \cite{hu2000,klaer} and other Bose-Einstein condensate DM theories have used smaller particle masses and obtained vortex cores on kpc scale \cite{zinner}. In our investigation we chose to create a one meter vortex core in the center of the galaxy, focusing on upholding all the constraints placed upon the parameters and the model in \cite{original} and \cite{Berezhiani2017}. Our main results from this part of the paper are as follows:
   \begin{enumerate}
      \item We can obtain a central one meter vortex core using the parameters $m\approx 1.1\times10^{-4}\,\rm{eV}$ and $\Lambda \approx 141\,\rm{MeV}$. This reproduces the zero-temperature condensate density profile of the original $m=1\,\rm{eV}$ and $\Lambda = 0.2\,\rm{meV}$ parameters. 
      \item All constraints of the model presented in \cite{original} and \cite{Berezhiani2017} seem to be upheld under these parameter changes.
      \item If the parameters are changed consistently in order to reproduce the same density profile, we will also obtain the same rotation curves. This means that observing rotation curves might not tell us about the values of $m$ and $\Lambda$, but could tell us about the relation between them.
      \item Too low particle masses, $< 10^{-10}\,\rm{eV}$, result in $\Lambda$-values beyond Planck mass scales and will in the more extreme cases result in overlapping vortices. 
   \end{enumerate}
In this work we have used the zero-temperature DM density profile as a reference when changing the parameters. We should point out that the DM in reality has some finite temperature which should be taken into account, and that the gravitational effect of the DM on the surrounding baryons also is omitted from our calculation of the density profile (based on \cite{original}). In \cite{Berezhiani2017} these effects are taken into account, which lead to the condensate radius estimate presented in equation \ref{eq:R_T}, as opposed to estimating the radius based on when the zero-temperature density profile hits zero. For our simple vortex analysis and parameter investigations we used the less complicated zero-temperature profile, but chose our parameters carefully so that the two different estimates gave similar results. In general, the more advanced density profile consideration results in smaller condensate components, which could affect the possibilities of observing the effects of vortices directly, for example through lensing. The more advanced condensate density profile of \cite{Berezhiani2017} also has a higher central density compared to the zero-temperature profile, which would result in slightly smaller vortex cores in the center, further motivating the search for parameters allowing for larger and more ``massive'' vortices.  

To test all model constraints we chose a specific vortex size to create, changing the DM particle mass to obtain the desired size. We then changed $\Lambda$ to reproduce the original density profile, as this one was based on the parameters used in \cite{original} and \cite{Berezhiani2017}. We have not tried to get as close as possible to the original density profile in the parameter variation, but tried to stay in the vicinity to ensure a realistic profile. The parameters we ended up with for creating a one meter vortex core in the center of the galaxy is quite different from the original values used, but seem to satisfy all constraints of the model. The new $m$ and $\Lambda$ values result in a very weak self-interaction for the DM particles, namely $\sigma/m\sim10^{-18}\,\rm{cm}^2/\rm{g}$. Observational constraints from galaxy cluster mergers on DM self-interaction give an upper limit of $\sigma/m\lesssim 0.5\,\rm{cm}^2/\rm{g}$ \cite{harvey}. This is, however, based on a calculation involving standard DM particles and could possibly change in a two-component picture. In addition, there is also the possibility that $\Lambda$ and $\sigma/m$ are related in some unknown way. This could limit the parameter space, as decreasing $m$ increases $\Lambda$ and decreases $\sigma/m$, stretching the two parameters in opposite extremes. 

A large portion of \cite{Berezhiani2017} is devoted to providing realistic rotation curves for two specific galaxies, using observational data of baryonic density profiles and detailed calculations for the SFDM profile. It is important that the model can provide sound rotation curve predictions, and it is also important that we keep this property when changing the parameters of the model. We therefore ran a simplified test using toy profiles for both the DM and baryonic components, and calculated rotation curves for four different parameter combinations. When picking an $m$-value and adjusting the $\Lambda$-value to reproduce the zero-temperature SFDM density profile of the original parameters, we obtain rotation curves that are all similar, as expected. If we choose a different $\Lambda$-value instead, the profile will look different, as we no longer operate with the same SFDM density profile, as illustrated in Figure \ref{fig:rot_c_vel}. From the figure we also see that the curves are not completely identical. This is most likely due to our method of estimating the $\Lambda$-values to reproduce the original density profile, and would be expected to disappear if enforcing a stricter matching criteria. This investigation also tells us that observing rotation curves might not give us a value estimate of $m$ and $\Lambda$, as there are many combinations that give the same result, but it could tell us about the relation between $m$ and $\Lambda$ instead. In \cite{Berezhiani2017} it is also discovered that the SFDM model predicts a slight rise in the rotation curve at large radii for high-surface-brightness (HSB) galaxies. The authors show, in their Figure 6, that keeping $m$ constant and varying $\Lambda$ gives different slopes, providing a possible observational identifier of the SFDM model and the parameter values. It was also found in \cite{zinner} that a vortex grid could leave an imprint on the galactic rotation curve. This was not further explored here as our vortices have a very low gravitational impact, combined with the fact that the aforementioned paper found the effect to vanish for vortex cores smaller than $5\;\rm{kpc}$. 

%In addition it would be interesting to further investigate if the full vortex grid in this framework could induce similar imprints on the rotation curve as seen in \cite{zinner}. This has not been done here as the results from the aforementioned paper shows that the effect vanishes for vortex cores sizes lower that $5\;\rm{kpc}$. 
%One may also think that the vortices could leave some imprint on the rotation curve, similar to what is discovered in \cite{zinner}. Based on our analysis here, however, this seems unlikely because of the small gravitational impact of the vortices in this framework.  

As a part of our parameter space investigation we also tested a few very low DM particle masses, as these can provide larger vortex cores. At $m\sim10^{-10}\,\rm{eV}$ we find that $\Lambda$-values close to the Planck mass scale is needed to reproduce the original zero-temperature SFDM density profile. Still using our constant density framework, this would give $\sim1000\rm{km}$ vortex cores in the galaxy center, and approximately ten times larger cores towards the edge of the condensate. The mass of all of the vortices in this scenario adds up to approximately one Earth mass, which is an extremely small fraction of the overall mass of a galaxy. Still, the constant density approach might be insufficient to estimate the effect of vortices at this scale, and it could also be the case that the overall grid formation have an effect on the baryons at a large scale. This should be investigated further. It should also be pointed out that the particles proposed for this SFDM model are described as ``axion-like''. This means that nothing more specific has been proposed than lightweight bosonic particles, another reason for why we wanted to look at masses different from the ones proposed in \cite{original} and \cite{Berezhiani2017}. 

In addition to a constant density case and a $m-\Lambda$ parameter space investigation we also considered the possibility of a vortex solution of the relativistic completion of the SFDM model presented in \cite{original}. By calculating the Euler-Lagrange equation of the full Lagrangian with respect to the modulus $\rho$, we arrived at a second-order nonlinear ODE. We have tried to solve this equation in various ways, and are yet to find a satisfying vortex solution. From our shooting method attempt is seems like the equation is highly unstable, and depends heavily on the initial guess of the parameters $\dot{R}_{\rm{init}}$ and $\Lambda_c$. For now, we have not let $m$ and $\Lambda$ vary during our attempts to solve the equation, and it could be possible that this would open up for some more stable regions in the $(m,\Lambda,\Lambda_c,\dot{R}_{\rm{init}})$ parameter space, where a vortex solution could be found. Another possibility is that the shooting method is unfit for our situation and that a more advanced and specialized method should be applied. This would have to be further investigated in the future. We are then left with a few different options. There could exist a vortex solution similar to what is expected in a regular Bose-Einstein condensate, as presented in \cite{pethick_smith_2008}, Section 9.2. There could exist a vortex solution that is different from what we would expect, but it could also be the case that the Lagrangian presented in \cite{original} does not have a vortex solution. There is also the possibility that a different approach should be taken to find a vortex solution than the one used in this paper, or that some of the assumptions used throughout the calculation do not hold. Based on the current analysis we can not strictly rule out any of the options, however, our results indicate that the possible vortex equation obtained here is highly unstable and require a fine tuning of various parameters to inch closer to something that could meet the expectations of a vortex profile. As the Lagrangian from \cite{original} was constructed with the goal of reproducing the MOND scalar action in the non-relativistic limit and MOND-regime, $\Lambda_c\ll|\phi|$, it is not certain that the existence of a vortex solution was considered when making the theory. In other words, it could be the case that there is no vortex solution. If so, it might be preferable to construct a new Lagrangian for the theory which encompasses this aspect of superfluidity. We should also address that the Lagrangian in equation \ref{eq:7.1} is made to reproduce the zero-temperature MOND scalar action, and does not include final temperature effects. Altering the Lagrangian to include this could then result in a different vortex equation with a possible solution. Also, \cite{mistele2021} has recently proposed a modified model where the roles of carrying the energy density of the superfluid and mediating the MOND-like force are split between two fields. This proposal could lead to a different vortex equation.

In summary, the vortices of the SFDM model are very small and separated by vast distances. The $m-\Lambda$ parameter space of the model is large, and smaller DM particle masses along with larger $\Lambda$-values result in larger and more gravitationally significant vortices. It is possible to alter the parameters in such a way that we keep the same condensate density profile, and by doing so, also the same rotation curves. Ideally, we would like to model the full vortex grid, without the assumption of rigid body rotation, to better understand how a full grid formation would impact surrounding baryons. We have so far not been able to find a plausible vortex solution for the Lagrangian presented in \cite{original}. All in all, any observationally detectable impact from the vortices on their surroundings seem improbable in the current framework. %There are, however, other observational markers that could help distinguish the SFDM model from other theories, such as the predicted slight raise of rotation curves for large radii or possibly the transition region between the SFDM model and a regular NFW profile at the edge of the DM condensate, as mentioned in \cite{Berezhiani2017}. In \cite{cai}, gravitational waves are also proposed as a way of exploring the parameter space of the SFDM model. 

\acknowledgments
This paper summarizes the results of a Masters Thesis which can be found at \url{https://www.duo.uio.no/handle/10852/69987} along with the full calculations and more in depth details regarding the overall work. Thank you to Robert Hagala for valuable input regarding the implementation of the shooting method, and to Benoit Famaey for an interesting discussion concerning the SFDM framework.

%\paragraph{Note added.} This is also a good position for notes added
%after the paper has been written.


\bibliographystyle{unsrt}
\bibliography{main_article.bib}
%\begin{thebibliography}{99}

%\bibitem{a}
%Author, \emph{Title}, \emph{J. Abbrev.} {\bf vol} (year) pg.

%\bibitem{b}
%Author, \emph{Title},
%arxiv:1234.5678.

%\bibitem{c}
%Author, \emph{Title},
%Publisher (year).

% Please avoid comments such as "For a review'', "For some examples",
% "and references therein" or move them in the text. In general,
% please leave only references in the bibliography and move all
% accessory text in footnotes.

% Also, please have only one work for each \bibitem.

%\end{thebibliography}
\end{document}